\documentclass[journal,draftclsnofoot,onecolumn,12pt,twoside]{IEEEtranTCOM}

\usepackage[T1]{fontenc}
\usepackage{cite}
\ifCLASSINFOpdf
\else
\fi
\usepackage{amsmath}
\usepackage[cmintegrals]{newtxmath}
\usepackage{subcaption}
\usepackage[version=4]{mhchem}
\usepackage{nicefrac}
\usepackage{xcolor}
\usepackage{mathtools}
\usepackage{array}
\DeclarePairedDelimiterX{\infdivx}[2]{(}{)}{%
  #1\;\delimsize\|\;#2%
}
\newcommand{\infdiv}{\infdivx}

\def\BibTeX{{\rm B\kern-.05em{\sc i\kern-.025em b}\kern-.08em
    T\kern-.1667em\lower.7ex\hbox{E}\kern-.125emX}}
\usepackage{breqn}
\usepackage{multirow}
\usepackage{longtable}

\begin{document}

\title{What Is the Trait d'Union between Retroactivity and Molecular Communication Performance Limits?}

\author{Francesca~Ratti,~\IEEEmembership{Graduate~Student~Member,~IEEE,}
        Maurizio~Magarini,~\IEEEmembership{Member,~IEEE,}
        and~Domitilla~Del~Vecchio,~\IEEEmembership{Fellow,~IEEE}
\thanks{F. Ratti is with the Department
of Information, Electronics, and Bioengineering, Politecnico di Milano, 20133 Milan, Italy and with the Mechanical Engineering Department, Massachusetts Institute of Technology, Cambridge, MA 02139, USA e-mail: fratti@mit.edu.}
\thanks{M. Magarini is with the Department
of Information, Electronics, and Bioengineering, Politecnico di Milano, 20133 Milan,
Italy e-mail: maurizio.magarini@polimi.it.}
\thanks{
D. Del Vecchio is with the Mechanical Engineering Department, Massachusetts Institute of Technology, Cambridge, MA 02139, USA e-mail:
ddv@mit.edu.}
}

\maketitle

\begin{abstract}
Information exchange is a critical process in all communication systems, including biological ones. The concept of retroactivity represents the loads that downstream modules apply to their upstream systems in biological circuits. This paper focuses on studying the impact of retroactivity on different biological signaling system models, which present analogies with well-known telecommunication systems. The mathematical analysis is performed both in the high and low molecular counts regime, by mean of the Chemical Master Equation and the Linear Noise Approximation, respectively.  The aim is to provide analytical tools to maximize the reliable information exchange for different biomolecular circuit models. Results highlight how, in general, retroactivity harms communication performance. This negative effect can be mitigated by adding to the signaling circuit an independent upstream system that connects with the same pool of downstream systems. 
\end{abstract}


\IEEEpeerreviewmaketitle

\section{Introduction}

\IEEEPARstart{M}{olecular} Communication (MC) is a field of research that has gained relevance in recent years. This emerging discipline is directly inspired by natural communications between cells in biology~\cite{akyildiz2008nanonetworks,akan2017fundamentals}. Characterizing  living  cells  from  an  information  and  communication theoretical perspective is one of the keys to understand the fundamentals of MC system engineering~\cite{forney1998modulation}. The development of  mathematical models for many of the phenomena occurring in an MC system has gained notice lately. In particular, great attention has been paid to the methods of propagation of molecules in the extracellular environment~\cite{gresho1985advection,chang2005physical,hundsdorfer2013numerical,llatser2013detection,mosayebi2014receivers}. A step forward has been that of studying the communication performance in biochemical circuits.
In fact, an open research problem in MC is that of finding optimal ways to transfer information in biochemical circuits.
Several works have investigated this problem recently.
For example, in~\cite{liu2015channel} the authors analyze the channel capacity of a biological system considering both diffusion-based channel and ligand-based receiver. Moreover, in~\cite{pierobon2013capacity} the authors provide a closed-form expression for the information capacity of an MC system with a noisy channel. In~\cite{awan2017improving} a different approach is taken, where the authors use enzymatic reaction cycles to improve the upper bound on the mutual information for a diffusion-based MC system. The work in~\cite{cheng2017capacity} presents an analysis of the channel capacity in diffusive MC by considering intersymbol interference from all the previous time slots and the channel transmission probability in each time slot.

Furthermore, there exist several works in the literature investigating parallelisms between well-known telecommunication and MC models and evaluating information exchange performance in different biological scenarios. For example, in~\cite{lu2016effect} the authors focus on the interference that is generated in the case of a broadcast channel, i.e., when the same transmitter (e.g., a cell) sends the same message simultaneously to multiple receivers (e.g., a number of cells). In~\cite{atakan2008molecular},
the authors present and develop models for the molecular multiple-access, broadcast, and relay channels in a MC system and
perform a numerical analysis on their capacity expressions. The authors of~\cite{liu2013multiple} study the capacity of multiple-access
channel that is affected by the parameters of the diffusive channel
and ligand-receptor binding mechanisms. 
The work in~\cite{rouzegar2017channel} presents a training-based channel impulse response estimation for diffusive multiple-input multiple-output (MIMO) channels.  
In~\cite{koo2016molecular}, a MIMO design for MC is proposed, where multiple molecular emitters are utilized at the transmitter and multiple molecular detectors are utilized at the receiver. 
Various diversity techniques for MIMO transmissions based on molecular diffusion are proposed in~\cite{meng2012mimo} to improve the communication performance in nanonetworks in the presence of multi-user interference. 

The common thread among the aforementioned works is the focus on the communication in the extracellular environment, where propagation of molecules plays a fundamental role in the communication performance. In this paper, we concentrate on biomolecular circuits whose communication is not prominently affected by propagation, and, in particular, we focus on isolating the impact of \emph{retroactivity} on the communication performance. Retroactivity is the effect that downstream systems receiving a signal apply to upstream ones sending the signal. It is an extension of the concept of loading and impedance to biomolecular systems~\cite{del2008modular}. 
Since then~\cite{del2008modular}, retroactivity has been studied in the contexts of control theory and systems biology, with the aim of characterizing and efficiently designing modular biomolecular circuits~\cite{saez2008automatic, del2009engineering, jayanthi2010retroactivity, ossareh2011long, anderson2011model, del2013control,sivakumar2013towards, pantoja2015retroactivity, bhaskaran2016effect}.

The back-propagated signal generated by the interconnected components plays a role not only in the design of biomolecular circuits but also in the communication performance of the upstream system. There exist a few works in the literature considering the effects of retroactivity on the information exchange in molecular circuits. The author of~\cite{awan2016effect} presents preliminary results on the impact of retroactivity on the communication performance of a diffusive MC system, composed of one transmitter and one receiver.  A review of retroactivity in different signaling systems and genetic circuits can be found in~\cite{mcbride2019effect}.

The main contribution of this paper consists in the analytical investigation of the impact of retroactivity on the information exchange for different MC system models, by making parallelisms with some well-known telecommunication ones. The ultimate objective is to lay foundations of how to efficiently maximize the reliable information exchange when considering different signaling circuits. In the following, we illustrate the results we achieved in this direction. Preliminary outcomes were presented in~\cite{ratti2020impact}.

The paper is organized as follows. In Sec.~\ref{sec: overview}, we summarize the structure of the paper, by giving an overview of the work. In Sec.~\ref{sec: preliminaries}, we introduce the theoretical concepts on which our work lays its foundations, and we contextualize them by explaining the preliminary assumptions we made. In Sec.~\ref{sec: systemModel} we present the biochemical system models on which we perform the communication performance evaluation. The subject of Sec.~\ref{sec: results} is the presentation and discussion of the results we obtain. Last, in Sec.~\ref{sec: conclusion} we conclude the paper.

\section{Overview}\label{sec: overview}
\begin{figure}[!t]
  \centering
 \includegraphics[width=\linewidth]{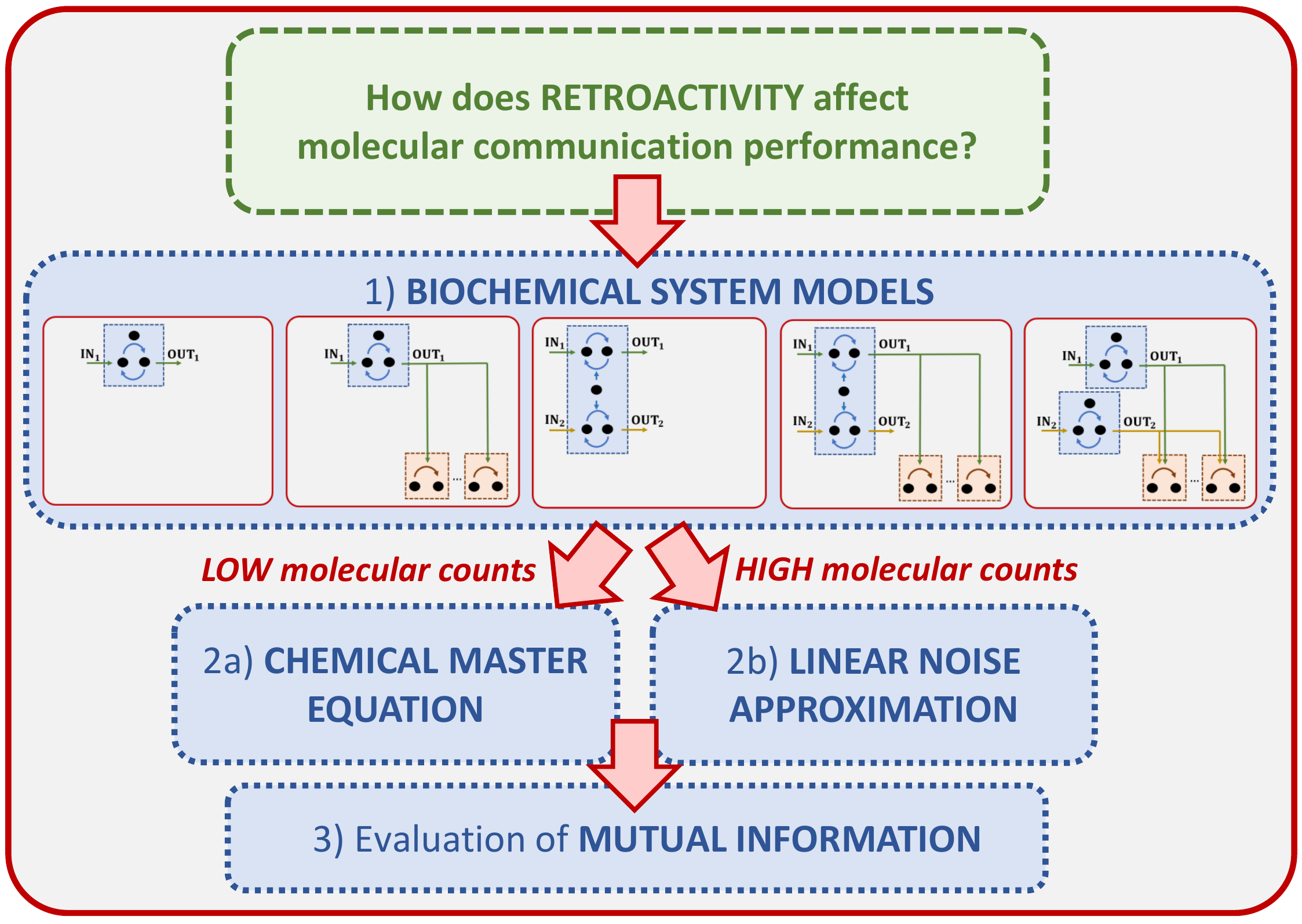}
  \caption{Overview of the structure of the paper. The biochemical system models are detailed in Fig.~\ref{fig: models}. The Chemical Master Equation (CME) is a system of differential equations that describes the rate of change of the probability of the system to be in any give state at time $t$~\cite{del2015biomolecular}. The Linear Noise Approximation (LNA) of the CME is a linear time-varying stochastic differential equation that allows a stochastic characterization of the evolution of a chemical reaction network, still maintaining scalability comparable to that of the deterministic models~\cite{van1992stochastic}.}
  \label{fig: overview}
\end{figure}

The fields of MC and of systems and synthetic biology have as a final goal that of understanding biological phenomena, and that of ``engineering'' them to improve upon their nature when possible or necessary, e.g., to combat diseases~\cite{chahibi2017molecular, veletic2019molecular}, or to enhance agricultural processes~\cite{van2010role, farsad2017novel}. We asked ourselves if the way a biological system is built can affect its communication performance, and, in particular, \emph{if and how retroactivity affects MC communication performance.} In fact, to the best of our knowledge, this question has not been answered in the literature yet. Awareness regarding the effect of retroactivity on communication performance is relevant both in the context of understanding and maximizing the information flow~\cite{kim2011modeling, kadloor2012molecular} and in that of designing biological circuits~\cite{alon2019introduction, grunberg2020modular}. The goal of this paper is that of filling this gap. 

Figure~\ref{fig: overview} visually summarizes the main steps we have taken. We consider five biochemical systems, which enable us to quantify the effect of retroactivity in a variety of scenarios. All of them present similarities with most of the common digital communication models. They will be detailed in Sec.~\ref{sec: systemModel}. 

We model the stochastic behavior of each biochemical system both in the low molecular counts regime via the Chemical Master Equation (CME)~\cite{del2015biomolecular}, and in the high molecular counts regime via the Linear Noise Approximation (LNA)~\cite{van1992stochastic}. Although the CME can describe systems in high molecular count regimes, its analysis in these cases is computationally demanding. Therefore, we opted for using the LNA in the high molecular count regime. 
In this way, we can compute a tractable analysis for the simpler models, and to make some speculations on the more complex ones.
Both the CME and the LNA are solved at \emph{steady state}. In fact, we consider the symbol completely received when the system reaches steady state, so that we avoid the effects of the transient behavior of the biomolecular circuit on the communication exchange performance, and we are able to isolate the contribution of the retroactivity. Furthermore, the steady state assumption intrinsically avoids the presence of intersymbol interference. The CME and LNA allow determining the probability mass functions (pmf) and the probability density functions (pdf), respectively, necessary for the evaluation of the mutual information between the input and the output of the system (Sec.~\ref{subsec: MIformulas}).

\section{Preliminaries}\label{sec: preliminaries}
The core of this work lays its foundations on concepts coming from different fields. In this section, we introduce the main theoretical background of this study.

\subsection{Retroactivity}

Each chemical reaction that, starting from a compound, produces another molecule can be regarded as an input/output system. Such input/output system can, in turn, be regarded as a module that, when connected with others, forms a more complex system. A fundamental issue that arises when interconnecting different components is how the process of transmitting a signal to a ``downstream'' module affects the dynamic state of the ``upstream'' subsystem sending the signal. In fact, upon interconnection, a signal that goes from the downstream to the upstream system is generated. This phenomenon is called \emph{retroactivity}, formally defined as the back action from the downstream system to the upstream one~\cite{del2008modular}.

\begin{figure}[!t]
  \centering
 \includegraphics[width=.8\linewidth]{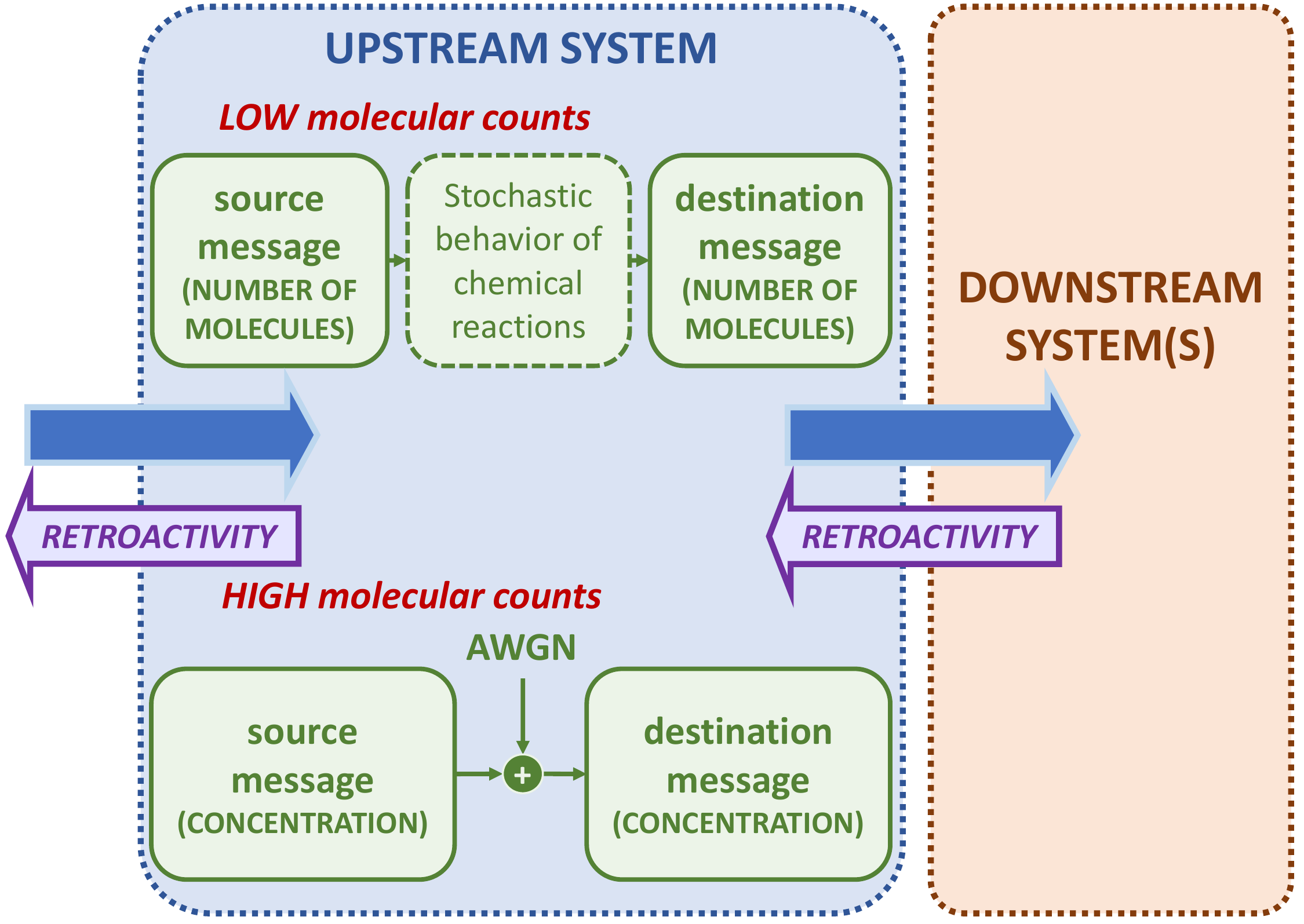}
  \caption{Upstream signaling system in the case of low and high molecular counts. The interconnection between the upstream and the downstream system generates the signal going from right to left in the figure (mauve arrow), that is the effect of retroactivity. The blue arrow from left to right represents the signal transmitted throughout the system.}
  \label{fig: retroactivity}
\end{figure}

In our paper, we consider different models of signaling circuits (see Sec.~\ref{sec: systemModel}). These are molecular circuits that, given external stimuli (inputs), through a series of chemical reactions, transform them to signals (outputs) that can control how cells respond to their environment. Then, we analyze to what extent retroactivity from downstream systems affects the communication performance between the input and the output of the upstream system.

Figure~\ref{fig: retroactivity} shows a general scheme of signal exchange between upstream and downstream systems. The blue arrows represent the signal passing through the input of the upstream system to the downstream system. The mauve arrows represent the effect of retroactivity, that is, the signal that goes from the downstream to the upstream system. We will perform the analysis for both the low and high molecular count regimes. In the figure, the upstream system describes the communication models representing both scenarios, which are detailed in the next sections.

\subsection{Communication system models}\label{subsec: comm system models}

It is well-known that the fundamental portions of any communication system are transmitter and receiver. The information signal is sent from the transmitter to the receiver via the communication channel. When the system is composed of only one transmitter and one receiver that exchange information on a communication channel, it is classified as \emph{Single-Input Single-Output} (SISO) transmission system~\cite{barry2012digital}. For instance, in digital communication, when both the transmitter and the receiver are composed by a single antenna, the system is categorized as SISO. In this scenario, the impediments that can degrade the performance are two: the noise affecting the communication channel and the intersymbol interference. In this work, we account for the noise affecting the communication due to stochasticity of the chemical reactions composing the communication system. 
However, it is not always true that the system is composed only of one transmitter and one receiver. In fact, there can be also multiple transmitters that, upon coordination, send information simultaneously to different receivers over the same communication channel. An example of this setup in wireless communication occurs in the case of multiple antennas both at the transmitter and at the receiver side. When this scenario occurs, we are in the presence of \emph{MIMO} transmission model~\cite{costa2010multiple}. In this case, the performance is affected also by the interference between the the different inputs.

Furthermore, communication systems can be classified based on different properties of the communication channel. In this paper, we will use the notions of \emph{Broadcast channel (BC)}, \emph{Multiple-Access Channel (MAC)}, \emph{Additive White Gaussian Noise (AWGN) channel}, and \emph{Z-channel}. These terms refer to different features of the channel. They are considered orthogonal definitions, except for BC and MAC that are mutually exclusive. In the following, we give a short framing for each of them. We talk about BC when one transmitter sends simultaneously over the channel the same or different messages to a number of receivers~\cite{barry2012digital}. The downlink from the base station to the portables is an example of BC in digital communication. On the contrary, the model composed by multiple isolated transmitters for which there is very little or no coordination and that communicate with the same receivers is called MAC~\cite{barry2012digital}. The term AWGN channel refers to the kind of noise that affects the communication, and it is one of the most studied types of noise in digital communication. In this paper, we will use the AWGN channel model in the case of high molecular counts, as highlighted in Fig.~\ref{fig: retroactivity}. In contrast, it is not possible to rely on a known statistical channel model in the low molecular counts regime. A Z-channel (binary asymmetric channel) is a channel with binary input and binary output, where each 0 is transmitted correctly, but each 1 has probability $P$ of being transmitted incorrectly as a 0, and probability $\left(1-P\right)$ of being transmitted correctly as a 1~\cite{mackay2003information}. 

\subsection{Information exchange quantification}\label{subsec: MIformulas}
In this paragraph we introduce the information theory background necessary for the evaluation of the communication performance carried out in this paper. Note that all the definitions in this paragraph and in the rest of the paper hold in a discrete domain, as it is in the low molecular count regime. The high molecular count regime case is solved considering continuous random variables, thus in all the formulas that follow the summation should be replaced with an integral.

The \emph{Mutual Information} (MI) of two random variables is a measure of the mutual dependence between them~\cite{cover1999elements}. In this paper, the MI is utilized as metric to quantify the information exchange between the input $X$ and the output $Y$ of the system. We calculate the MI as~\cite{cover1999elements}
\begin{equation}\label{MI}
    I\left(X,Y\right) = H\left(X\right)-H\left(X\mid Y\right),
\end{equation}
where $H\left(X\right)$ and $H\left(X\mid Y\right)$ are the entropy of the input and the entropy of the input given the output, respectively. Assuming $X$ and $Y$ to be discrete, these entropies can be calculated with the well-known formulas~\cite{cover1999elements}
\begin{equation}\label{HX}
    H(X) = -\sum_{j=0}^{M-1} P\left(x_j\right)\log P\left(x_j\right)\nonumber,
\end{equation}
and
\begin{align}\label{HXY}
    H\left(X\mid Y\right) & \hspace{-.05cm}=\hspace{-.05cm} -\sum_{i=0}^{M-1}\sum_{j=0}^{N-1} P\left(y_i\right)P\left(x_j\mid y_i\right)\log P\left(x_j \mid y_i\right)\nonumber,
\end{align}
where $M$ and $N$ are the sample spaces, and $P\left(X\right)$ and $P\left(Y\right)$ the pmfs of $X$ and $Y$, respectively. Note that in this paper we calculate the MI between $X$ and $Y$ at different instants of time. In fact, supposing $X$ the input of our communication system, we consider its pmf at $t_0$, i.e., the beginning of the symbol transmission. Conversely, the pmf of the output $Y$ is evaluated at steady state $t_s$, as previously mentioned. In this paragraph we avoid the use of the subscripts $t_0$ and $t_s$ to make the formulas more readable.

Another key concept in communication is that of channel capacity, i.e., the maximum of reliable information that can be sent over a communication channel.
In particular, we consider an upper and a lower bound on the channel capacity. The definition of the bounds is given starting from $P\left(X\right)$, $P\left(Y\right)$, and the conditional pmfs.

\subsubsection{The lower bound}
starting from~\eqref{MI}, the channel capacity is defined as~\cite{cover1999elements}
\begin{equation}\label{capacity}
   C = \max_{P\left(X\right)}I\left(X,Y\right),
\end{equation}
that is, the maximum of the reliable information that can be transmitted over a communication channel without memory.

It is possible to notice that, for any input distribution $P\left(X\right)$, ~\eqref{capacity} leads to a natural lower bound on the capacity
\begin{equation}\label{Formula: lower bound}
    C \ge I\left(X,Y\right).
\end{equation}

\subsubsection{The upper bound}

the Kullback-Leibler divergence ($D_{KL}$) leads to the definition of MI~\cite{csiszar2011information}
\begin{equation}\label{MIdual}
    I\left(X,Y\right) = D_{KL}\infdiv{P\left(X,Y\right)}{P\left(X\right)P\left(Y\right)}.
\end{equation}
Furthermore, from the definition of $D_{KL}$, it is possible to write
\begin{equation}\label{DKL}
   D_{KL}\infdiv{P(Y\hspace{-.05cm}\mid\hspace{-.05cm} x)}{P\left(Y\right)} \hspace{-.05cm}= \hspace{-.05cm}\sum_{y\in Y}P\left(y\mid x\right)\log \left( \frac{P\left(y\mid x\right)}{P\left(y\right)}\right). 
\end{equation}

In turn, we can rewrite~\eqref{MIdual} as
\begin{align}\label{dkl_new}
   I\left(X,Y\right) 
   &= \sum_{y\in Y}P\left(y\right) D_{KL}\infdiv{P\left(X\mid y\right)}{P\left(X\right)}.
\end{align}

From~\eqref{dkl_new} we can derive a dual formula to express the capacity for amplitude-constrained channels~\cite{csiszar2011information}
\begin{equation}
    C = \min_{P\left(Y\right)}\max_{x\in X}D_{KL}\infdiv{P\left(Y\mid x\right)}{P\left(Y\right)}.
\end{equation}
Therefore, every choice of a distribution $P\left(Y\right)$ leads to an upper bound on the channel capacity~\cite{lapidoth2003capacity}
\begin{equation}\label{Formula: upper bound}
   C \le \max_{x\in X}D_{KL}\infdiv{P\left(Y\mid x\right)}{P\left(Y\right)}.  
\end{equation}

We will also use the well-known formula of the Z-channel capacity~\cite{mackay2003information}, that starting from the conditional pmf $P\left(X = 1 \mid Y = 0\right) = P$, is given by the equation
\begin{equation}\label{ZchannelCapacity}
    C = \log \left(1+\left(1-P\right)P^{\nicefrac{P}{\left(1-P\right)}} \right),
\end{equation}
and that of the AWGN channel capacity~\cite{cover1999elements}, supposing a Gaussian output distribution $P\left(Y\right)\sim \mathcal{N}\left(\mu_Y,\sigma_Y^2\right)$,
\begin{equation}\label{AWGNchannelCapacity}
    C = \frac{1}{2}\log \left(1+\frac{\sigma_Y^2}{\sigma_{noise}^2}\right),
\end{equation}
where $\sigma_{noise}^2$ is the variance of the Gaussian-distributed noise.

\subsection{Stochastic models for biochemical systems}\label{subsec: CME,LNA}
The previous paragraph shows that to evaluate~\eqref{MI},~\eqref{Formula: lower bound}, and~\eqref{Formula: upper bound}, the marginal pmfs (pdfs in the continuous case) $P\left(X\right)$, $P\left(Y\right)$, and the conditional pmfs $P\left(Y \mid X\right)$, $P\left(X \mid Y\right)$ are required. Thus, a stochastic modeling to approximate the behavior of the considered systems at steady state is needed. Several modelings have been proposed in the literature~\cite{hougen1943chemical, gillespie1977exact, van1992stochastic, risken1996fokker, gillespie2000chemical}. As anticipated in Sec.~\ref{sec: overview}, in this paper we model the behavior of the system in the low molecular count regime via the CME~\cite{gillespie1977exact, del2015biomolecular}, while we use the LNA~\cite{van1992stochastic} under the assumption of high molecular counts.
\subsubsection{The Chemical Master Equation} the CME  describes the rate of change of the probability of a molecular microstate in a chemical reaction system. For a system with N chemical species, a microstate represents the number of molecules present for each species in the system at a given time. Thus, it requires the enumeration of all the possible microstates of the system (i.e., each possible microscopic configuration). In this work, as previously mentioned, we consider steady state, that is, we set to zero the rate of change of the probability in the CME.

The motivations behind the choice of the CME to model the behavior of the system are two. First, we consider the accuracy of the results worth the complexity of the calculations. Furthermore, the CME does not take into account any stochastic effects  other than the ones due to intrinsic noise in the chemical reactions. In this way, we can selectively quantify the effect of retroactivity on information exchange for different biomolecular system models. The major drawback of the CME is the computational complexity in computing the probability of being in each microstate at time $t$, which increases exponentially as the number of molecules and reactions in the system increases. For this reason, we use the CME only in the low molecular counts regime, and we rely on the LNA for studying the behavior of systems with a higher number of molecules.

\subsubsection{The Linear Noise Approximation} The LNA lays its foundations on the central limit theorem from probability theory~\cite{montgomery2014applied}, which states that, when independent random variables are added, their properly normalized sum tends toward a normal distribution, even if the original variables themselves are not normally distributed. This is why the LNA is valid only in the case of high molecular counts. Thus, when applying the LNA to model the stochastic behavior of a biochemical circuit, the resulting distributions of the species concentrations of the system in time are Gaussian. When using the LNA, the channel is assumed AWGN with variance $\sigma_{noise}^2 = 1$, hence simplifying the communication model. The mean of the  distribution is obtained by the reaction rate equations of the system~\cite{del2015biomolecular}. In our case, since we work at steady state, these differentials equations are solved for the equilibrium. The steady state covariance matrix of the system is computed by solving the Lyapunov equation~\cite{del2015biomolecular}. As it can be noticed in the upstream system of Fig.~\ref{fig: retroactivity}, in the case of the LNA model, the unit of measure of the biochemical species is concentration rather than number of molecules as in the CME. Thus, the number of molecules in the system is to be divided by the reaction volume. 

\begin{figure}
\centering
\begin{tabular}{cccc}
\includegraphics[width=0.3\textwidth]{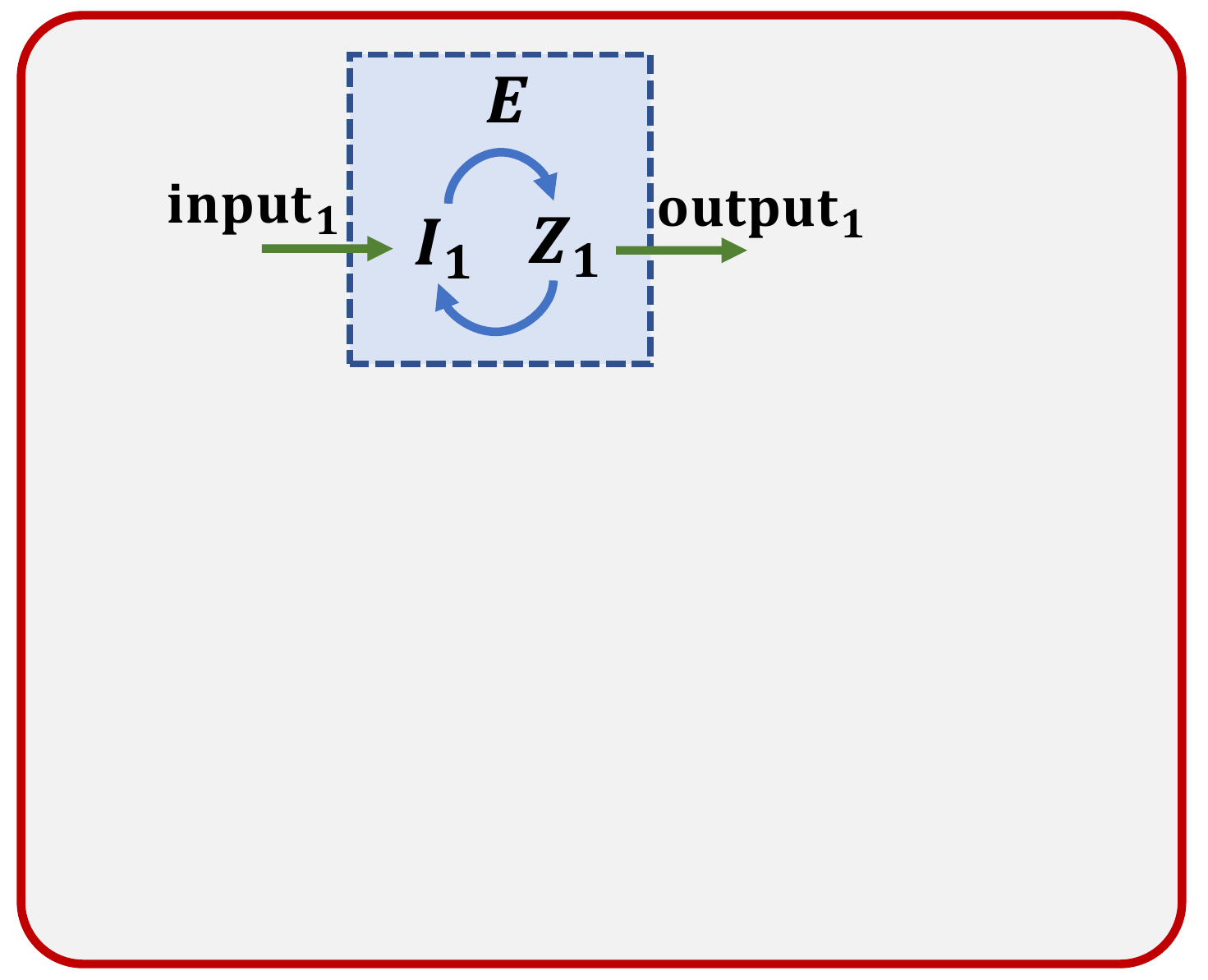} &
\includegraphics[width=0.3\textwidth]{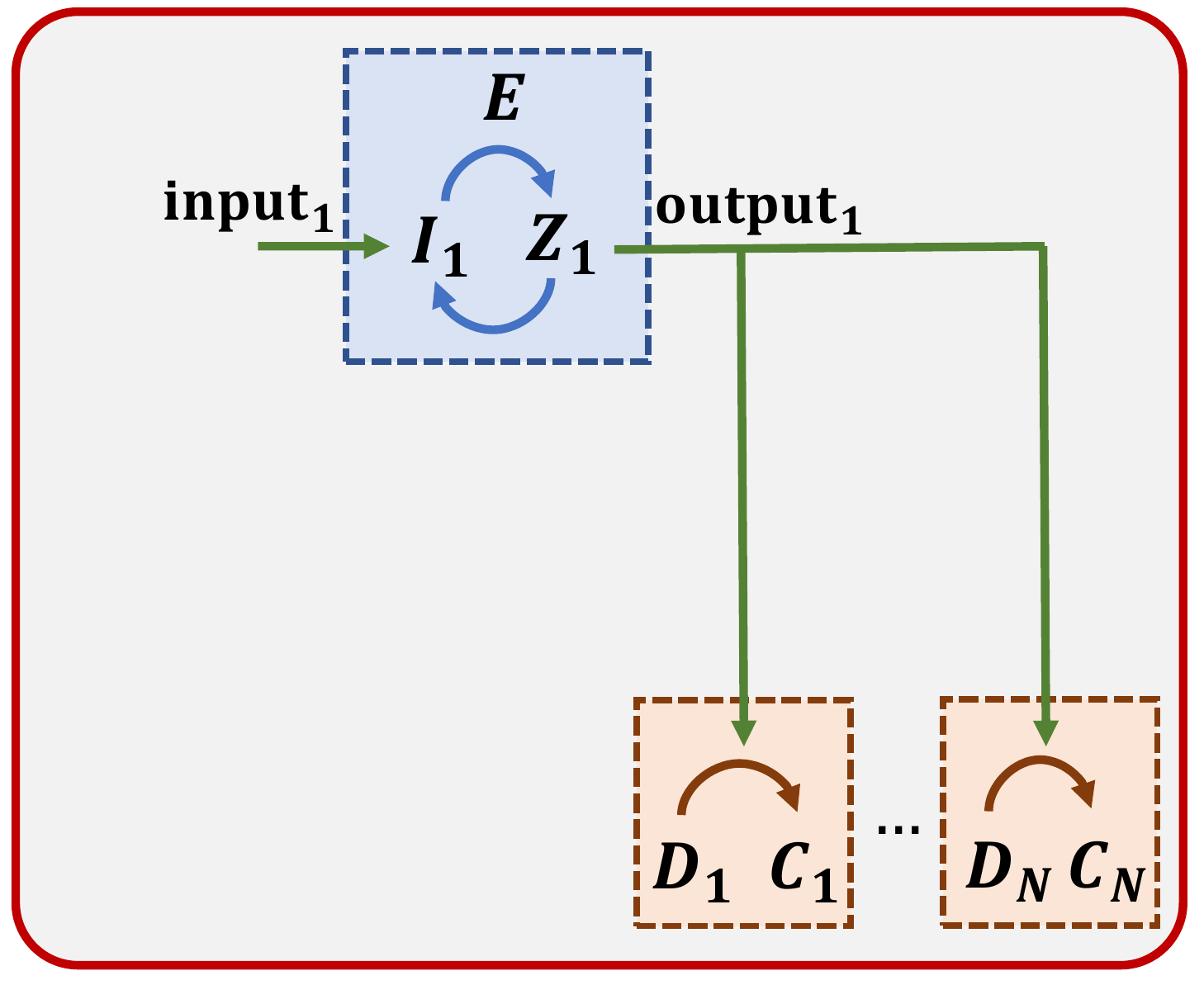} &
\includegraphics[width=0.3\textwidth]{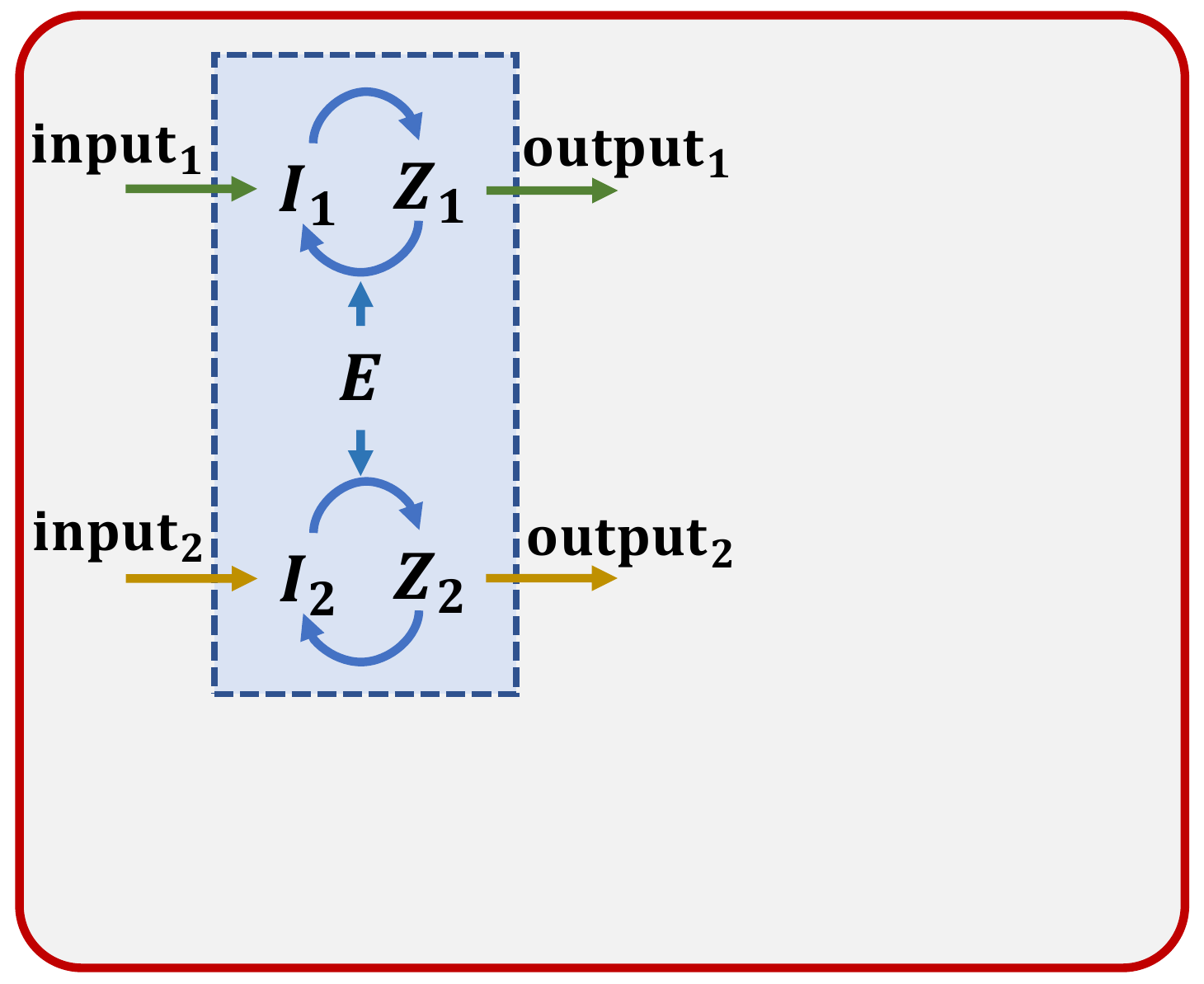} \\
\text{(a)}  & \text{(b)} & \text{(c)}  \\[6pt]
\end{tabular}
\begin{tabular}{cccc}
\includegraphics[width=0.3\textwidth]{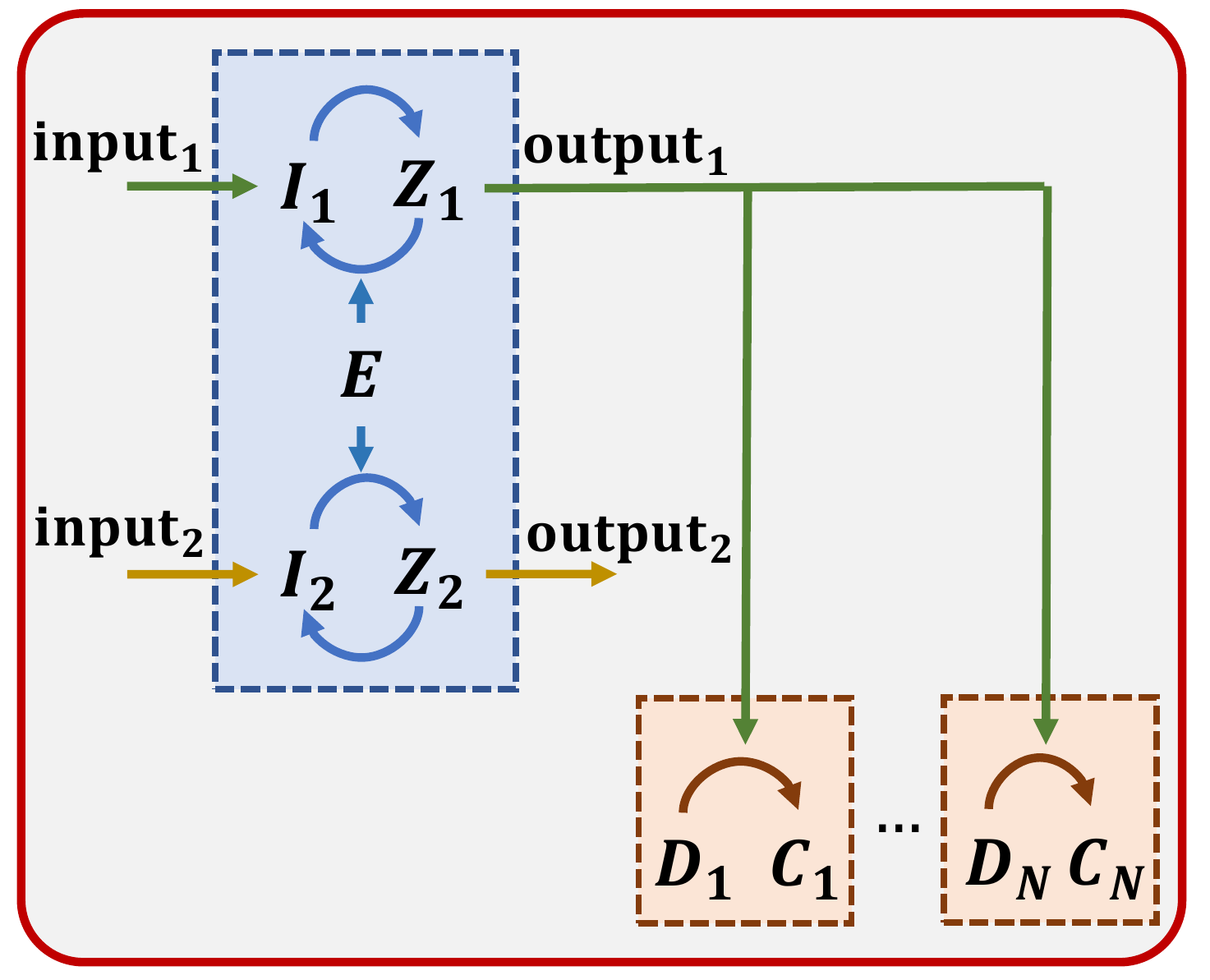} &
\includegraphics[width=0.3\textwidth]{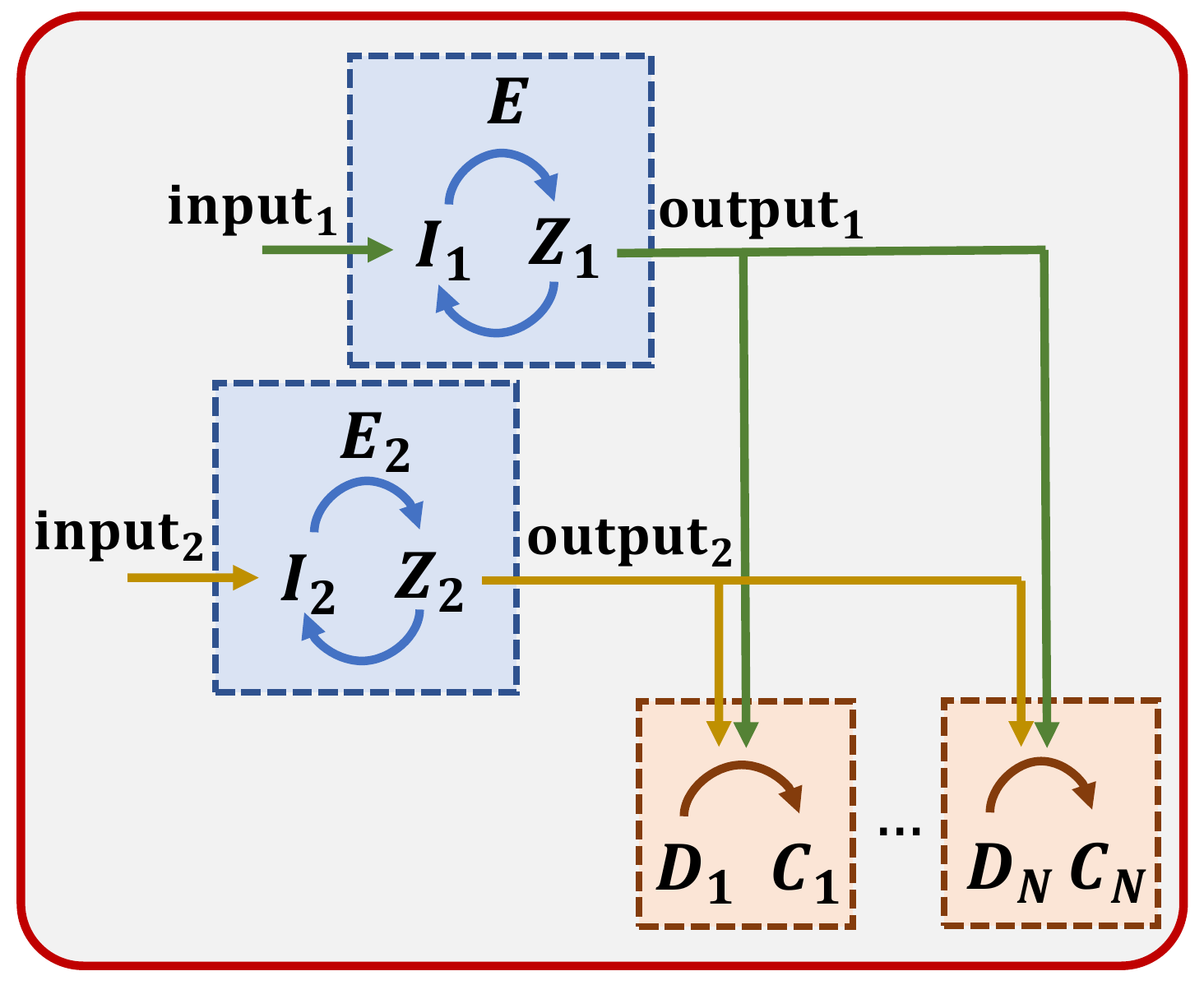} \\
\text{(d)}  & \text{(e)}  \\[6pt]
\end{tabular}
\caption{ Diagrams of the signaling systems analyzed in the paper: (a)~isolated SISO, (b)~SISO with $N$ downstream targets, (c)~isolated MIMO, (d)~MIMO with $N$ downstream targets, (e)~two isolated SISO with MAC.}
\label{fig: models}
\end{figure}

\section{System model}\label{sec: systemModel}
In this section, we present the chemical reaction models analyzed in the paper. More specifically, we consider the five signaling system models shown in Figure~\ref{fig: models}. Figure~\ref{fig: models}a represents an isolated signaling system, which is not connected to any downstream target. In digital communication, this could be viewed as a \emph{SISO system}, where the input is $\text{I}_1$, and the consequent output is $\text{Z}_1$. Then, in Fig.~\ref{fig: models}b, the same isolated system is connected through $\text{Z}_1$ to $N$ chemical circuits. Thus, in Fig.~\ref{fig: models}b, the system of Fig.~\ref{fig: models}a becomes an upstream system directly linked to $N$ downstream ones. This model could be viewed as a \emph{BC}, where the output of the upstream system $\text{Z}_1$ is the same message sent to different receivers, i.e., the downstream systems. The third signaling system (Fig.~\ref{fig: models}c) models the same upstream system in the presence of a second circuit that shares the same enzyme E. We model the second circuit as a replica of the first one. This recalls the \emph{MIMO} setup typical of the telecommunication systems, where the multiple inputs are $\text{I}_1$ and $\text{I}_2$ and the outputs are $\text{Z}_1$ and $\text{Z}_2$, respectively. The fourth considered signaling system is illustrated in Fig.~\ref{fig: models}d. It is a MIMO model in which the output of the first upstream system $\text{Z}_1$ is connected to $N$ downstream targets. The last model considers two isolated SISO upstream systems that can connect to the same $N$ downstream systems. Note that only one of the two upstream systems can connect with a specific downstream system at a time. This means that when both upstream systems are sending information, the number of downstream systems available to receive the message $\text{Z}_1$ is $Q \le N$. This recalls the MAC of digital communication, where the two upstream systems are the different transmitters that do not coordinate, and the downstream systems are the receivers that communicate with both the transmitters.

The two-step enzymatic reaction representing the upstream system can be described as
\begin{equation}
\begin{array}{lcc}
    \ce{$\text{I}_1$ + E <=>[$\text{k}_0$] $\text{M}_1$}\\
    \ce{$\text{M}_1$ ->[$\text{c}_1$] E + $\text{Z}_1$}\\ \ce{$\text{Z}_1$ ->[$\text{c}_2$] $\text{I}_1$}\nonumber,
\end{array}
\end{equation}
 with conservation laws $\text{E}_{\text{tot}} = \text{E}+\text{M}_1$, and $\text{I}_{\text{tot}_1} = \text{I}_1+\text{M}_1+\text{Z}_1$, where $\text{I}_1$ is the protein (input message) that binds to an enzyme E to form the complex $\text{M}_1$, that in turn is transformed in the output protein $\text{Z}_1$~\cite{del2015biomolecular}. The coefficients $\text{c}_1$ and $\text{c}_2$ are the catalytic rates of the unidirectional chemical reactions, and $\text{k}_0 = \nicefrac{\text{k}_0^-}{\text{k}_0^+}$ is the so called dissociation constant of the reversible binding reaction, $\text{k}_0^+$ and $\text{k}_0^-$ being the association and dissociation rate constants~\cite{del2015biomolecular}. Note that the last reaction makes the system cyclic. This is a major discrepancy with respect to traditional communication models, which are unidirectional. The secondary upstream circuit in the last three models is composed of an analogous two-step enzymatic reaction, with input $\text{I}_2$ and output $\text{Z}_2$, with conservation law $\text{I}_{\text{tot}_2} = \text{I}_2+\text{M}_2+\text{Z}_2$. Thus, the conservation law that preserves the total amount of enzyme present in the system becomes $\text{E}_{\text{tot}} = \text{E}+\text{M}_1+\text{M}_2$ for the models in Fig.~\ref{fig: models}c and~\ref{fig: models}d. The model in Fig.~\ref{fig: models}e is characterized by two separate conservation laws that preserve the quantity of the enzymes $\text{E}$ and $\text{E}_2$, that are, respectively,  $\text{E}_{\text{tot}_1} = \text{E}+\text{M}_1$ and $\text{E}_{\text{tot}_2} = \text{E}_2+\text{M}_2$.
 
 Each $j$th downstream system is composed of one reversible reaction
\begin{equation}
 \ce{Z_1 + \text{D}_j <=>[$\text{k}_{3_j}$] \text{C}_j} \nonumber 
\end{equation}
 with conservation law $\text{D}_{\text{tot}_j} = \text{D}_j + \text{C}_j$ Here, the output of the upstream system $Z_1$ (and also $Z_2$ in Fig.~\ref{fig: models}e) binds with the DNA D to form the complex C. Note that for the model in Fig.~\ref{fig: models}e, the conservation law becomes $\text{D}_{\text{tot}_j} = \text{D}_j + \text{C}_{j_1} + \text{C}_{j_2}$, where $\text{C}_{j_1}$ is the complex formed by the binding of $\text{Z}_1$ with $\text{D}$ and $\text{C}_{j_2}$ the one coming from the binding of $\text{Z}_2$ with $\text{D}$.  The coefficient $\text{k}_{3_j}$ is the dissociation constant of the reversible binding reactions and is equal to $\nicefrac{\text{k}_{3_j}^-}{\text{k}_{3_j}^+}$, where $\text{k}_3^+$ and $\text{k}_3^-$ are the association and dissociation rate constants, respectively.
 The presence of the downstream systems modifies the conservation laws of $\text{I}_{\text{tot}_1}$ and $\text{I}_{\text{tot}_2}$. These become $\text{I}_{\text{tot}_1} = \text{I}_1+\text{M}_1+\text{Z}_1+\sum_{j=1}^Q\text{C}_j$ and $\text{I}_{\text{tot}_2} = \text{I}_2+\text{M}_2+\text{Z}_2+\sum_{j=1}^{N-Q}\text{C}_j$, where $Q$ is the number of upstream systems connected to the first upstream system, that in the second and in the fourth model is equal to $N$, while in the last one we have $0 \le Q \le N$.
Thus, supposing no $\text{M}_1$, $\text{M}_2$, $\text{Z}_1$, $\text{Z}_2$, $\text{C}_j$ present in the system before $t_0$, we can write $\text{I}_{\text{tot}_1}\left(t_0\right) = \text{I}_{1}\left(t_0\right)$ and $\text{I}_{\text{tot}_2}\left(t_0\right) = \text{I}_{2}\left(t_0\right)$.

\section{Results: Communication Performance Evaluation}\label{sec: results}
In this section, we present the results obtained supposing a low and a high molecular count regime.
Note that, due to the complexity of the calculations, not for all the case studies we determine an analytical formula of the MI. Nevertheless, we discuss whether retroactivity causes a reduction of the communication performance in all the considered scenarios.

\subsection{Low molecular counts}
For all the signaling system models, we make the same hypotheses on the values of $\text{E}_{\text{tot}}$, $\text{D}_{\text{tot}_j}$ and on the number of input symbols $n_{\text{I}_1}$, $n_{\text{I}_2}$, and their corresponding $\text{I}_1\left(t_0\right)$, $\text{I}_2\left(t_0\right)$ to compute the steady state solution of the CME. We choose $n_{\text{I}_1} = n_{\text{I}_2} = 2$. This means that both the upstream systems have two available input symbols, one composed by 0 molecules of I ($\text{I}\left(t_0\right) = 0$, no transmission) while the second composed by one molecule of I ($\text{I}\left(t_0\right) = 1$). This corresponds to a Concentration On-Off Keying modulation~\cite{kuran2011modulation}.
This choice allows the explicit enumeration of all the microstates of the system, thus the analytical investigation of the role of retroactivity in the information exchange between $\text{I}_1\left(t_0\right)$ and $\text{Z}_1\left(t_s\right)$, where $t_s$ stands for steady state time.
Accordingly, we also set   $\text{E}_{\text{tot}}$,  $\text{D}_{\text{tot}_j} = 1$. 

From~\eqref{MI}, by solving the CME to obtain the necessary pmfs and by substituting $X$ with $\text{I}_1\left(t_0\right)$ and $Y$ with $\text{Z}_1\left(t_s\right)$, it is possible to show (see the derivations in the Appendix) that the formula of the MI we obtain in the five considered cases is the same, corresponding to
\begin{align}\label{formula: MI}
    &I\left(\text{I}_1\left(t_0\right), \text{Z}_1\left(t_s\right)\right) = \log_e\left(P_{0_1}^{\left(-P_{0_1}\right)}P_{1_1}^{\left(-P_{1_1}\right)}\right)\\
    &-\log_e\left(\left(\left(\frac{P_{0_1}}{P_{0_1}+AP_{1_1}}\right)^{\left(-P_{0_1}\right)}\right)\cdot\left(\left(\frac{AP_{1_1}}{P_{0_1}+AP_{1_1}}\right)^{\left(-AP_{1_1}\right)}\right)\right)\nonumber
\end{align}
nat/symbol, where $P_{0_1}$ and $P_{1_1}$ are the probabilities of the two possible input symbols $\text{I}_1\left(t_0\right) = 0$ and $\text{I}_1\left(t_0\right) = 1$. The term $A$ is a constant depending on the coefficients of the chemical reactions of the upstream and downstream systems, and on the probabilities $P_{0_2}$ and $P_{1_2}$ of the input symbols of the second upstream system in the two MIMO models. 

In the first case, when the model is composed by a single isolated SISO signaling system (Fig.~\ref{fig: models}a), we obtain 
\begin{equation}\label{eq: A0}
    A = A_0 = \frac{\left(1+\textbf{k}_0\right)\text{c}_2}{\text{c}_1+\left(1+\textbf{k}_0\right)\text{c}_2}, 
\end{equation}
where $\textbf{k}_0 = \nicefrac{\left(\text{k}_0^-\Omega\right)}{\text{k}_0^+}$, $\Omega$ being the volume in which the reactions take place, as detailed in the appendix.
This shows that the probability of the output $\text{Z}_1\left(t_s\right)$ is affected by the presence of the reaction $\ce{Z_1 ->[$\text{c}_2$] I_1}$ that transforms the output protein $\text{Z}_1$ in the input $\text{I}_1$, not allowing the complete consumption of $\text{I}_1$ at steady state.

The term $A$ in the second system (Fig.~\ref{fig: models}b) captures the impact of retroactivity on the information exchange due to $N$ downstream systems. In this case, the formula of $A$ is affected by the coefficients of the reactions connecting the upstream and the downstream systems. We obtain 
\begin{equation}\label{eq: aN}
    A=A_N = 1-\frac{\text{c}_1}{\left(1+\textbf{k}_0\right)\text{c}_2+\left(1+\sum_{j=1}^N\left(\nicefrac{1}{\textbf{k}_{3_j}}\right)\right)\text{c}_1},
\end{equation}
where $\textbf{k}_{3_j} = \nicefrac{\left(\text{k}_{3_j}^-\Omega\right)}{\text{k}_{3_j}^+}$.
 Note that if we remove the downstream systems effect $\sum_{j=1}^N\left(\nicefrac{1}{\textbf{k}_{3_j}}\right)$, we return to $A_0$. This system captures the impact of retroactivity on the information exchange performance in a generic biomolecular signaling BC. If we assume that all the $N$ downstream systems have the same dissociation constant $\textbf{k}_{3}$, we can rewrite $A_N = 1-\frac{\text{c}_1}{\left(1+\textbf{k}_{0}\right)\text{c}_2+\left(1+\nicefrac{N}{\textbf{k}_{{3}}}\right)\text{c}_1}$. We observe that, for $N \to \infty$, $A_N$ is equal to 1. By substituting $A = 1$ in~\eqref{formula: MI}, we obtain that $I\left(\text{I}_1\left(t_0\right), \text{Z}_1\left(t_s\right)\right) = 0$, i.e., $\text{I}_1\left(t_0\right)$, $\text{Z}_1\left(t_s\right)$ are completely independent. This is a coherent result if we note  that, for infinite downstream systems, the probability of the number of free molecules of $\text{Z}_1\left(t_s\right)$ being different from 0 tends to 0, independently of the value of  $\text{I}_1\left(t_0\right)$.

We also note that, for $\text{c}_2 = 0$, i.e., if the upstream system is not cyclic, $A_0$ becomes 0, leading to $I\left(\text{I}_1\left(t_0\right), \text{Z}_1\left(t_s\right)\right) = \log_e\left(P_{0_1}^{\left(-P_{0_1}\right)}P_{1_1}^{\left(-P_{1_1}\right)}\right)$. This corresponds to having the MI equal to the entropy of $\text{I}_1\left(t_0\right)$. Thus, in this case $\text{I}_1\left(t_0\right)$ and $\text{Z}_1\left(t_s\right)$ are fully dependent. On the contrary, removing the cycle in the upstream system results in $A_N = 1-\frac{\text{c}_1}{\left(1+\sum_{j =1}^N\left(\nicefrac{1}{\textbf{k}_{3_j}}\right)\right)\text{c}_1}$. This is because the load given by the connected downstream systems still has an impact on the information exchange between $\text{I}_1$ and $\text{Z}_1$ by subtracting free $\text{Z}_1$ (i.e., the output message) from the system environment. If $\text{c}_2 \to \infty$ we get $A = A_0 = A_N = 1$, thus the MI becomes 0, meaning that the input and the output are completely independent. In fact, for $\text{c}_2 \to \infty$, the produced $\text{Z}_1$ is immediately transformed into $\text{I}_1$, so that the number of free molecules $\text{Z}_1\left(t_s\right)$ is equal to 0 for all values of $\text{I}_1\left(t_0\right)$. In general, the higher is $A$, the lower is the communication exchange performance measured via~\eqref{formula: MI}.

The formula of $A$ for the third model (Fig.~\ref{fig: models}c) is influenced by the presence of the second upstream system
\begin{equation}\label{eq: a0_MIMO}
    A = A_{0,\ \text{MIMO}} = A_0 P_{0_2} + B P_{1_2},
\end{equation}
where $A_0$ comes from~\eqref{eq: A0}, and $B$ is a constant depending on the chemical reaction rates as follows. Supposing all equal catalytic rates $\text{c}_1 = \text{c}_2 = \text{c}$ and supposing binding/unbinding rates $\text{k}_0^+, \text{k}_0^- \gg \text{c}$, we can write $B$ as
\begin{equation}\label{eq: B}
    B = \frac{3\left(\textbf{k}_0^+\right)^2 + \left(\text{k}_0^-\right)^2 + 4\text{k}_0^-\textbf{k}_0^+ + 3\text{k}_0^-\text{c} + 6\textbf{k}_0^+\text{c}}{5\left(\textbf{k}_0^+\right)^2 + \left(\text{k}_0^-\right)^2 + 5\text{k}_0^-\textbf{k}_0^+ + 3\text{k}_0^-\text{c} + 8\textbf{k}_0^+\text{c}},
\end{equation}
where $\textbf{k}_0^+ = \nicefrac{\text{k}_0^+}{\Omega}$ (see appendix).
If $P_{1_2} = 0$ and by consequence $P_{0_2} = 1$, we obtain $A_{0,\ \text{MIMO}} = A_{0}$, thus we return to the single isolated SISO model. 
We demonstrated via Wolfram Mathematica\textsuperscript{\textregistered} software that $A_0 P_{0_2} + B P_{1_2} \ge A_0$ is always true, for every value of $P_{0_2}$ and $P_{1_2}$, with the constraint that their sum should always be 1 and they cannot assume negative values. Thus, since $A_{0,\ \text{MIMO}}$ is always greater than $A_0$, the MI in~\eqref{formula: MI} is always lower in the case of isolated MIMO rather than in the case of isolated SISO. As in digital communication, also in MC, the communication performance of a MIMO system is worse than the ones of a SISO system given the same boundary conditions.

Likewise, the formula of $A$ for the fourth system (Fig.~\ref{fig: models}d) becomes
\begin{equation}\label{eq: aN_MIMO}
    A = A_{N,\ \text{MIMO}} = A_N P_{0_2} + G P_{1_2}
\end{equation}
where $A_N$ is~\eqref{eq: aN} and $G$ depends on the rates of the chemical reactions. The analytical formula of $G$ cannot be easily expressed in closed form, although it is worth noticing that it depends not only on the reaction rates of the upstream, but also on the binding/unbinding rates of the downstream systems. This is a key difference with respect to $B$~\eqref{eq: B}. Under some assumptions, i.e., all equal (un)binding rates $\text{k}^+, \text{k}^-$, all equal catalytic rates c, and $\text{k}^+, \text{k}^- \gg \text{c}$, we obtain via Wolfram Mathematica\textsuperscript{\textregistered} software $G > B$ and $A_N > A_0$. From these, it follows $A_{N,\ \text{MIMO}} > A_{0,\ \text{MIMO}}~\eqref{eq: a0_MIMO}$, that is, when the output $\text{Z}_1$ of the MIMO upstream system get connected with $N$ downstream systems, the MI lowers. 

In this case, when $P_{1_2} = 0$, we return to the second model, i.e., the SISO upstream system and BC. Notwithstanding this, we would like to understand if $A_{N,\ \text{MIMO}}$ is always greater than $A_{N}$, i.e., if the presence of a second upstream system lowers the MI. To do that, we set the inequality $A_{N,\ \text{MIMO}} \ge A_{N}$, and we solve it with respect to $G$. Specifically
\begin{align}\label{eq: G}
    &A_N P_{0_2} + G P_{1_2} \ge A_N \nonumber\\
    \Rightarrow \ &A_N \left(P_{0_2} -1\right) - G A_N \left(P_{0_2} -1\right) \ge 0 \nonumber\\
    \Rightarrow \ &G \ge A_N.
\end{align}
From~\eqref{eq: G}, we observe that it would be theoretically possible that for some values of $G$ (i.e., for $G < A_N$), $I\left(\text{I}_1\left(t_0\right), \text{Z}_1\left(t_s\right)\right)$ of the system in Fig.~\ref{fig: models}d becomes higher than the one in Fig.~\ref{fig: models}b. This would imply that a MIMO upstream system binding with the same number $N$ of downstream as the single SISO upstream can mitigate the impact of retroactivity on $I\left(\text{I}_1\left(t_0\right), \text{Z}_1\left(t_s\right)\right)$. 

 \begin{figure}
    \centering
    \begin{subfigure}[t]{0.65\linewidth}
        \centering
        \includegraphics[width=1\linewidth]{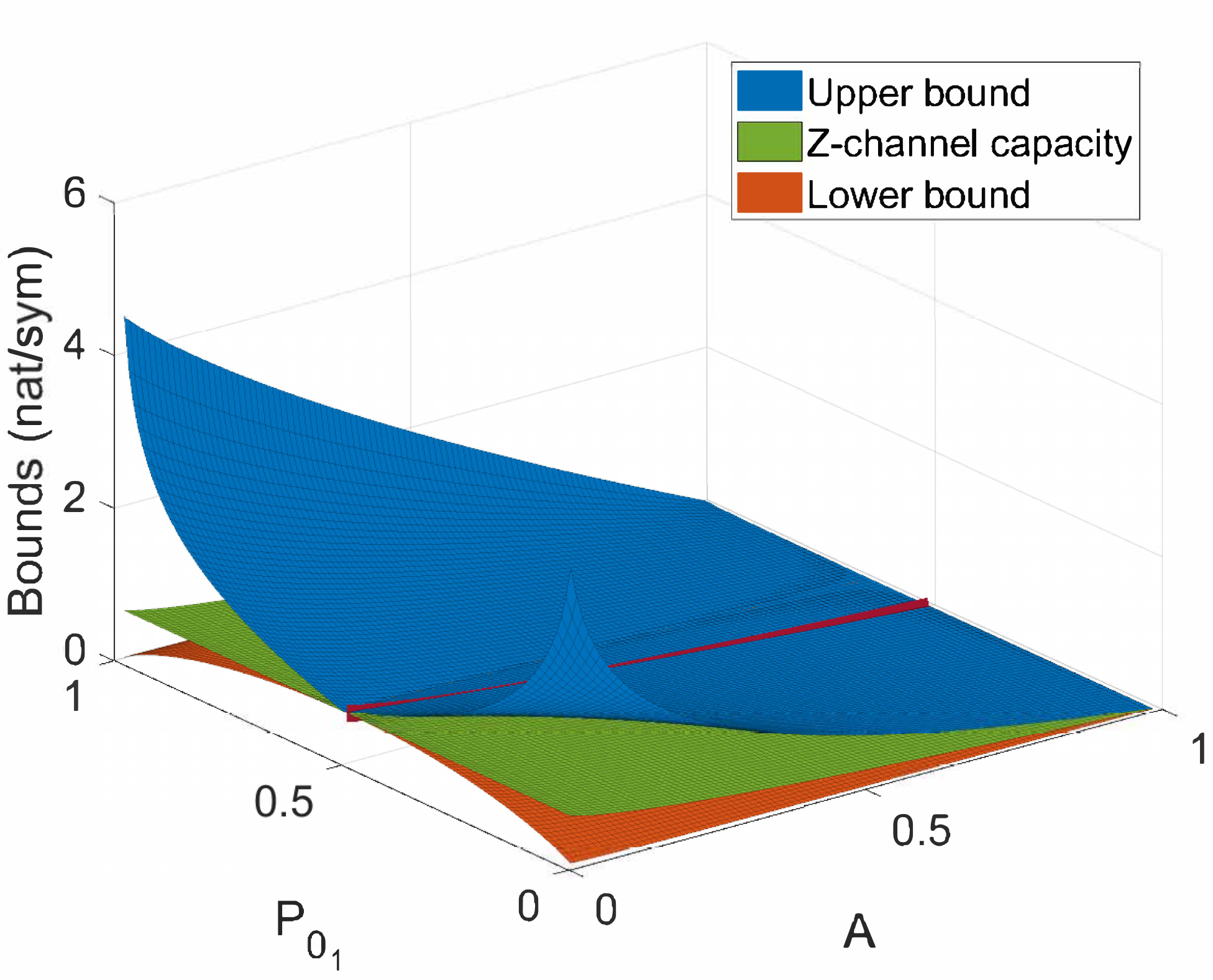}
        \caption{}
        \label{fig: 3Dbounds}
    \end{subfigure}
    \begin{subfigure}[t]{0.55\linewidth}
        \centering
        \includegraphics[width=1\linewidth]{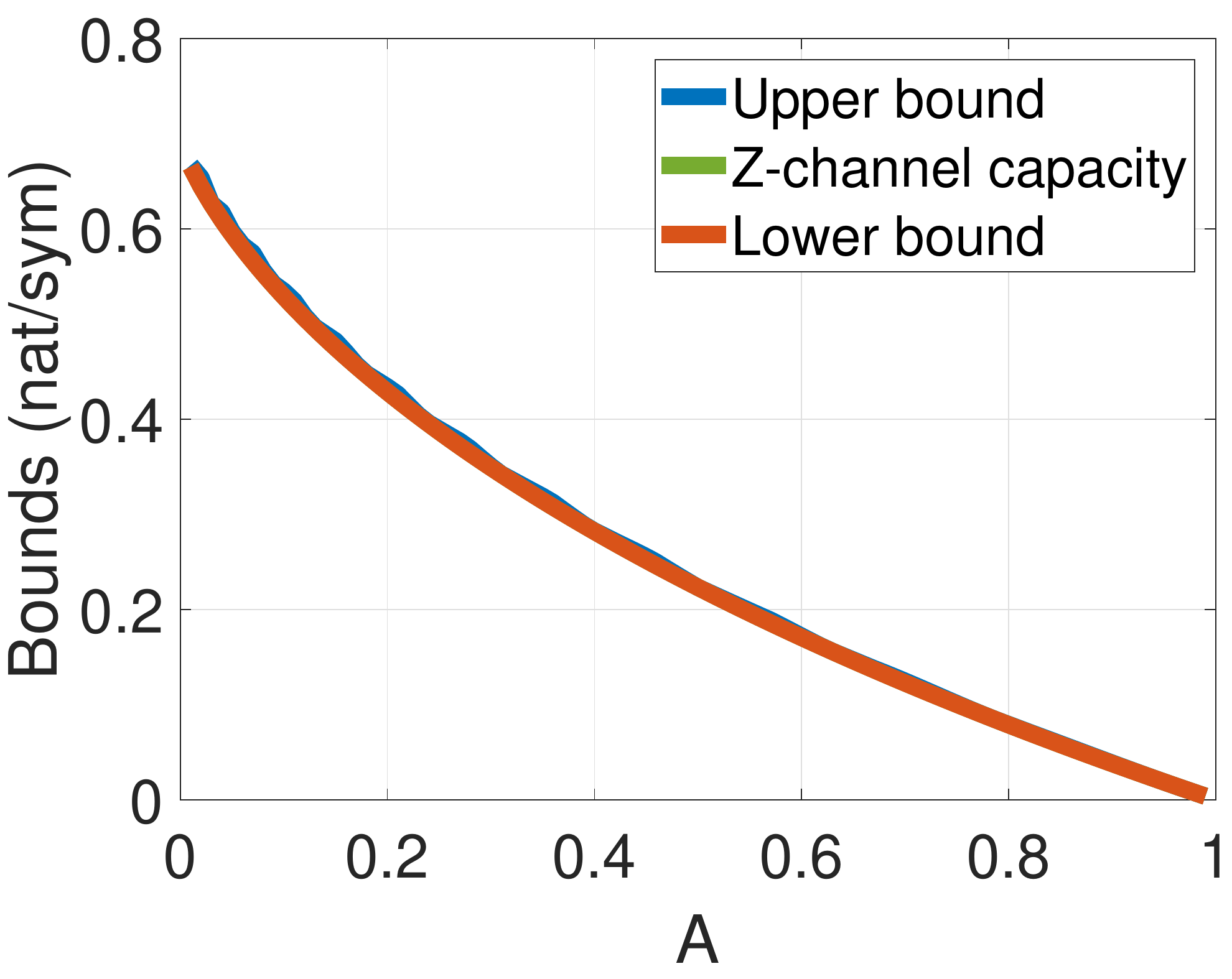}
        \caption{}
       \label{fig: 2Dbounds}
    \end{subfigure}
    \caption{(a)~Upper, lower bounds on the channel capacity, and Z-channel capacity for the four signaling systems, by varying $P_{0_1}$ and $A$. (b)~Channel capacity bounds and Z-Channel capacity with respect to the valid range of $A$ values.}
    \label{fig: F2}

\end{figure}

The expression of $A$ in the model of Fig.~\ref{fig: models}e is noteworthy. In fact, although there are two upstream systems in the environment, they are completely independent, meaning that the presence of the second does not have any influence on the communication performance of the first one. Since the two systems share the binding with the same $N$ downstream systems, $A$ becomes
\begin{equation}\label{eq: AMAC}
    A = A_{Q,\ \text{MAC}} = 1-\frac{\text{c}_1}{\left(1+\textbf{k}_0\right)\text{c}_2+\left(1+\sum_{j=1}^Q\left(\nicefrac{1}{\textbf{k}_{3_j}}\right)\right)\text{c}_1}.
\end{equation}
 The number of downstream systems binding with the second upstream is always complementary to $Q$, and it is equal to $N-Q$. Supposing that the two upstream systems emit the same number of molecules, i.e., $\text{Z}_1$, $\text{Z}_2$, then on average $Q = \nicefrac{N}{2}$. If $Q = 0$, then~\eqref{eq: AMAC} becomes~\eqref{eq: A0}, while in the opposite case, when $Q = N$, then $A_{Q,\ \text{MAC}} = A_N$~\eqref{eq: aN}. From this, we note that the fifth model represents an improvement in relation to the communication performance with respect to the model of Fig.~\ref{fig: models}b. In fact, since the second upstream system binds with some of the available downstream systems, the retroactivity observed at the first upstream system is lowered.

Note that regardless of the specific signaling system, each of these cases can be traced back to a Z-channel. In fact, the joint pmf $P\left(\text{I}_1\left(t_0\right), \text{Z}_1\left(t_s\right)\right)$ is for all the five models
\begin{equation}\label{formula: jointPMF}
    P\left(\text{I}_1\left(t_0\right), \text{Z}_1\left(t_s\right)\right) = \begin{bmatrix} P_{0_1} & 0 \\ AP_{1_1} & \left(1-A\right)P_{1_1}\end{bmatrix},
\end{equation}
where the rows represent the 0 and 1 possible values of $\text{I}_1\left(t_0\right)$ and the columns that of $\text{Z}_1\left(t_s\right)$ (see appendix for the exact derivation).
From~\eqref{formula: jointPMF}, it is clear that if $\text{I}_1\left(t_0\right) = 0$, then $\text{Z}_1\left(t_s\right)$ is certainly equal to 0, while if $\text{I}_1\left(t_0\right) = 1$, then $\text{Z}_1\left(t_s\right)$ is 1 with a certain probability $\left(1-A\right)$ and 0 with probability $A$. This is exactly the Z-channel behavior described in Sec.~\ref{subsec: comm system models}. 
For this reason, we validate our MI formula~\eqref{formula: MI}, by plotting the Z-channel capacity~\eqref{ZchannelCapacity}, with $P$ that in our case is to be substituted with $A$, and by varying the values of $A$ and of $P_{0_1}$. Note that, since both $A$ and $P_{0_1}$ are probabilities, the range of valid values that they can assume is included between 0 and 1. Furthermore, we plot on the same graph the MI we obtained~\eqref{formula: MI}. By doing this, as it can be observed from~\eqref{Formula: lower bound}, we are also finding a lower bound on the channel capacity for each $\left(A, P_{0_1}\right)$ and, for each $A$, what is the value of $P_{0_1}$ that maximize it, thus the value of the channel capacity. By plotting also the upper bound as described in~\eqref{Formula: upper bound} in 3D space, we note that the minimum of the upper bound, the Z-channel capacity, and the maximum of the lower bound for each value of $A$ coincide (red line). Figure~\ref{fig: 3Dbounds} is the visualization of what has just been described. Thanks to this 3D plotting, we are able to isolate the channel capacity value for each value of $A$ (Fig.~\ref{fig: 2Dbounds}), that, as for each value of the MI, decreases as $A$ increases, confirming our theoretical results.

\subsection{High molecular counts}
As mentioned in Sec.~\ref{subsec: CME,LNA}, the LNA describes the system through an ordinary differential equation and a linear (time-varying) stochastic differential equation~\cite{van1992stochastic}. To mitigate the analytical complexity, when only a single upstream system is present, we reduce it to a one-step cyclic enzymatic reaction
\begin{equation}
\begin{array}{lcc}
    \ce{$\text{I}_1$ + E ->[$\text{c}_1$] E + $\text{Z}_1$}\\ \ce{$\text{Z}_1$ ->[$\text{c}_2$] $\text{I}_1$}\nonumber,
\end{array}
\end{equation}
with conservation law $\text{I}_{\text{tot}_1} = \text{I}_1 + \text{Z}_1 + \sum_{j=1}^N\text{C}_j$. 
Thus, in the case of isolated SISO system, the differential equation characterizing the variation of the mean of $Z_1$ with time can be written as
\begin{equation}\label{eq: diffUP}
    \frac{d\text{Z}_1}{dt} = \text{c}_1\text{E}\left(\text{I}_{\text{tot}_1}-\text{Z}_1\right) - \text{c}_2\text{Z}_1,
\end{equation}
where $\sum_{j=1}^N\text{C}_j$ is equal to 0 (no downstream systems). Then, the mean of the Gaussian distribution characterizing $Z_1\left(t_s\right)$ can be found easily by setting~\eqref{eq: diffUP} equal to 0 and by solving the resulting equation with respect to ${Z}_1$
\begin{equation}
    \mu_0 = \frac{\text{c}_1\text{E}\text{I}_{\text{tot}_1}}{\text{c}_2+\text{c}_1\text{E}}.
\end{equation}
The variance $\sigma_0^2$ of the same distribution is obtained by solving the Lyapunov equation
\begin{align}\label{eq: var0}
   2\left(-\text{c}_1\text{E}-\text{c}_2\right)\sigma_0^2 &= -\frac{1}{\Omega}\left(\text{c}_1\text{E}\left(\text{I}_{\text{tot}_1}-\frac{\text{c}_1\text{E}\text{I}_{\text{tot}_1}}{\text{c}_2+\text{c}_1\text{E}}\right)+\text{c}_2\frac{\text{c}_1\text{E}\text{I}_{\text{tot}_1}}{\text{c}_2+\text{c}_1\text{E}}\right)\nonumber\\
   \Rightarrow \sigma_0^2 &= \frac{\text{c}_1\text{E}\text{I}_{\text{tot}_1}\text{c}_2}{\Omega\left(\text{c}_1\text{E}+\text{c}_2\right)^2},
\end{align}
where $\Omega$ is the volume of the environment (i.e., usually a cell).
Then, remembering that the LNA supposes $\sigma_{noise}^2 = 1$, it is straightforward deriving the channel capacity by substituting~\eqref{eq: var0} in~\eqref{AWGNchannelCapacity}.

Supposing equal dissociation constant $\text{k}_3$ for all the downstream systems, their characterizing differential equation is
\begin{equation}
    \frac{d\text{C}_j}{dt} = \text{k}_3^+\text{Z}_1\text{D}_j-\text{k}_3^-\text{C}_j = \text{k}_3^+\text{Z}_1\left(\text{D}_{\text{tot}_j}-\text{C}_j\right)-\text{k}_3^-\text{C}_j,
\end{equation}
where $\text{Z}_1$ is to be substituted with $\text{Z}_2$ when the second upstream system connects with the $j$th downstream one. Thus, the steady state value of $\text{C}_j$ when connected to the first upstream system is
\begin{equation}\label{eq: cjLNA}
    \text{C}_j = \frac{\text{Z}_1\text{D}_{\text{tot}_j}}{\text{k}_3+\text{Z}_1}.
\end{equation}

Then, for the second model (SISO with $N$ downstream systems, Fig.~\ref{fig: models}b) we obtain 
\begin{dmath}\label{eq: zNLNA}
    \frac{d\text{Z}_1}{dt} = \text{c}_1\text{E}\left(\text{I}_{\text{tot}_1}-\text{Z}_1-\sum_{j = 1}^N\text{C}_j\right) + \text{k}_3^-\sum_{j=1}^N\text{C}_j 
    - \text{c}_2\text{Z}_1 - \text{k}_3^+\text{Z}_1\sum_{j=1}^N\left(\text{D}_{\text{tot}_j}-\text{C}_j\right).
\end{dmath}
In order to get the mean value of $\text{Z}\left(t_s\right)$ distribution, we should substitute~\eqref{eq: cjLNA} in~\eqref{eq: zNLNA}. Then, supposing that the $N$ downstream systems have equal conservation law $\text{D}_{\text{tot}} = \text{D}+\text{C}$, we obtain $\mu_N = \mu_0$. This means that in the high molecular count regime the mean of the $\text{Z}\left(t_s\right)$ distribution is not impacted by the presence of the downstream systems. Due to the higher computational complexity, we do not provide the formula of the variance $\sigma_N^2$ in this case. However, to determine it one would have to solve the Lyapunov equation for this system with respect to $\sigma_N^2$. 

A two-step enzymatic reaction for the upstream systems is intrinsic of MIMO models. Thus, we reuse the chemical reactions modeling of the low molecular counts regime. For this reason, the system of differential equations characterizing the third model (Fig.~\ref{fig: models}c) is
\begin{equation}\label{eq: 3sysLNA}
    \begin{dcases}
    \frac{d\text{Z}_1}{dt} = \text{c}_1\text{M}_1 -\text{c}_2\text{Z}_1\\
    \frac{d\text{M}_1}{dt} = \text{k}_0^+\text{E}\text{I}_{1} - \left(\text{k}_0^-+\text{c}_1\right)\text{M}_1\\
    \frac{d\text{Z}_2}{\text{dt}} = \text{c}_1\text{M}_2 - \text{c}_2\text{Z}_2\\
    \frac{d\text{M}_2}{dt} = \text{k}_0^+\text{E}\text{I}_{2} - \left(\text{k}_0^-+\text{c}_1\right)\text{M}_2.
    \end{dcases}
\end{equation}
From the conservation laws, $\text{I}_1 = \text{I}_{\text{tot}_1} - \text{M}_1 - \text{Z}_1$, $\text{I}_2 = \text{I}_{\text{tot}_2}- \text{M}_2-\text{Z}_2$, and $\text{E} = \text{E}_{\text{tot}}-\text{M}_1-\text{M}_2$.

Similarly, the system in Fig.~\ref{fig: models}d can be written as
\begin{equation}\label{eq: 4sysLNA}
    \begin{dcases}
    \frac{d\text{Z}_1}{dt} = \text{c}_1\text{M}_1+ \text{k}_3^-\sum_{j=1}^N\text{C}_j -\text{Z}_1\left(\text{c}_2 +\text{k}_3^+\sum_{j=1}^N\left(\text{D}_{\text{tot}_j}-\text{C}_j\right)\right)\\
    \frac{d\text{M}_1}{dt} = \text{k}_0^+\left(\text{E}_{\text{tot}}-\text{M}_1-\text{M}_2\right)\left(\text{I}_{\text{tot}_1} - \text{M}_1 - \text{Z}_1-\sum_{j = 1}^N\text{C}_j\right) - \left(\text{k}_0^-+\text{c}_1\right)\text{M}_1\\
    \frac{d\text{Z}_2}{dt} = \text{c}_1\text{M}_2 - \text{c}_2\text{Z}_2\\
    \frac{d\text{M}_2}{dt} = \text{k}_0^+\left(\text{E}_{\text{tot}}-\text{M}_1-\text{M}_2\right)\left(\text{I}_{\text{tot}_2} - \text{M}_2 - \text{Z}_2\right) - \left(\text{k}_0^-+\text{c}_1\right)\text{M}_2,
    \end{dcases}
\end{equation}
where $\text{C}_j$ is to be substituted with~\eqref{eq: cjLNA} in the first two differential equations.

Both~\eqref{eq: 3sysLNA} and~\eqref{eq: 4sysLNA}, after being set equal to 0, are to be solved with respect to $\text{Z}_1$ to obtain the mean value of the Gaussian-distributed $\text{Z}\left(t_s\right)$. We do not report here these results and the Lyapunov equation that would give the variance in these cases, due to their analytical complexity. 

The differential equation characterizing $\text{Z}_1$ of the fifth system (Fig.~\ref{fig: models}e) is analogous to~\eqref{eq: zNLNA}, considering the number of downstream systems $0 \le Q \le N$ and remembering $\text{D}_{\text{tot}} = \text{D} + \text{C}_{j_1}+ \text{C}_{j_2}$. The same analytical complexity applies for the calculation of the variance. 

\section{Conclusions and Discussion}\label{sec: conclusion}
In this paper, we evaluate the effect of retroactivity on communication performance of different signaling systems. More specifically, we considered five different biomolecular circuits having analogies with some well-known digital communication system models. For each of them, we evaluated the analytical formula of the MI between the input at time 0 (i.e., the beginning of the transmission of a symbol) and the output evaluated at steady state. The steady state assumption implies the absence of intersymbol interference and the mitigation of the transient behavior of the system. 

Our results show that retroactivity has a negative impact on the communication performance, by reducing the MI as the number of downstream systems increases. We note also how the effect of retroactivity can be mitigated by the presence of a second independent upstream system, that connects with the same pool of downstream systems. The same behavior occurs for all the five systems both in the high and in the low molecular count regime.

The analytical computation of the Gaussian distributed output in the high molecular count regime for the remaining cases, as well as a biological interpretation of the $A$ constant in the low molecular count regime are left for future work.
Nevertheless, we can make some hypotheses on the analytical results in the high molecular counts. For the biochemical model in Fig.~\ref{fig: models}b, given that in the low molecular count regime the MI of this system is lower than the one in the SISO case, and given~\eqref{AWGNchannelCapacity}, we can speculate that $\sigma_N^2 = \sigma_0^2 - K$, where $K$ is a positive constant. If true, this would imply a lower channel capacity value than in the first scenario (Fig.~\ref{fig: models}a). Regarding the model of Fig.~\ref{fig: models}e, since the number of downstream systems connected to $Z_1$ is less or equal than the one of the second case, we speculate that in this case the variance $\sigma_{Q,\ \text{MAC}}^2$ would be greater than $\sigma_N^2$.

\appendix  
This appendix is devoted to the derivation of the expression of the MI~\eqref{formula: MI}. In all the five considered scenarios, the first step to perform is that of recalling the marginal distribution of the input 
\begin{equation}
    P\left(\text{I}_1\left(t_0\right)\right) = \begin{bmatrix}P_{0_1} \\ P_{1_1}\end{bmatrix},
\end{equation}
and that of evaluating $P\left(\text{I}_1\left(t_0\right) \mid \text{Z}_1\left(t_s\right)\right)$. To do this, it is necessary to solve the CME. The first step consists in enumerating all the possible microstates of the biomolecular system. Clearly, for the same number of molecules, as the number of species present in the system increases, also the number of possible microstates becomes higher. In our work, we consider $n_{\text{I}_1} = n_{\text{I}_2} = 2$, thus  $\text{I}_{\text{tot}_1}\le 1$ and $\text{I}_{\text{tot}_2}\le 1$. Furthermore, we set $\text{E}_{\text{tot}} = \text{D}_{\text{tot}_j} = 1$. The following tables~\ref{table: n1},~\ref{table: n2} resume all the possible microstates $q_i$ for each considered system when $\text{I}_{\text{tot}_1} = 0$, $\text{I}_{\text{tot}_2} = 1$ and $\text{I}_{\text{tot}_1} = 1$, $\text{I}_{\text{tot}_2} = 1$ respectively. Note that we consider $\text{I}_{\text{tot}_2} = 1$ because we are interested in the value of the MI in the presence of the molecule emitted by the second upstream system. For simplicity, we suppose to have only one downstream system, i.e., $j = 1$.
\begin{center}
\small
\begin{longtable}{
|c|c|c|c|c|c|c|c|c|c|c|c|c| } 
\caption{Microstates of the systems when $\text{I}_{\text{tot}_1} = 0$, $\text{I}_{\text{tot}_2} = 1$. \label{table: n1}}\\
\hline
System model & State & $\text{M}_1$ & $\text{I}_1$ & $\text{Z}_1$ & $\text{C}_1$ & $\text{M}_2$ & $\text{I}_2$ & $\text{Z}_2$ & $\text{C}_2$ & E & $\text{E}_2$ & D \\
\hline
\multirow{1}{7em}{Isolated SISO} & $q_1$ & 0 & 0 & 0 & 0 & 0 & 0 & 0 & 0 & 1 & 0 & 0 \\ 
\hline
\multirow{1}{7em}{SISO + downst.} & $q_1$ & 0 & 0 & 0 & 0 & 0 & 0 & 0 & 0 & 1 & 0 & 1\\ 
\hline
\multirow{3}{7em}{Isolated MIMO} & $q_1$ & 0 & 0 & 0 & 0 & 0 & 0 & 1 & 0 & 1 & 0 & 0 \\ 
& $q_2$ & 0 & 0 & 0 & 0 & 0 & 1 & 0 & 0 & 1 & 0 & 0 \\ 
& $q_3$ & 0 & 0 & 0 & 0 & 1 & 0 & 0 & 0 & 0 & 0 & 0 \\
\hline
\multirow{3}{7em}{MIMO with 1 downstream target}
& $q_1$ & 0 & 0 & 0 & 0 & 0 & 0 & 1 & 0 & 1 & 0 & 1 \\ 
& $q_2$ & 0 & 0 & 0 & 0 & 0 & 1 & 0 & 0 & 1 & 0 & 1 \\
& $q_3$ & 0 & 0 & 0 & 0 & 1 & 0 & 0 & 0 & 0 & 0 & 1 \\
\hline
\multirow{4}{7em}{Two isolated SISO with MAC} & $q_1$ & 0 & 0 & 0 & 0 & 0 & 0 & 0 & 1 & 0 & 1 & 0 \\ 
& $q_2$ & 0 & 0 & 0 & 0 & 0 & 0 & 1 & 0 & 0 & 1 & 1 \\ 
& $q_3$ & 0 & 0 & 0 & 0 & 0 & 1 & 0 & 0 & 0 & 1 & 1 \\
& $q_4$ & 0 & 0 & 0 & 0 & 1 & 0 & 0 & 0 & 0 & 0 & 1 \\
\hline
\end{longtable}
\end{center}

\begin{center}
\small
\begin{longtable}{
|c|c|c|c|c|c|c|c|c|c|c|c|c| } 
\caption{Microstates of the systems when $\text{I}_{\text{tot}_1} = 1$, $\text{I}_{\text{tot}_2} = 1$. \label{table: n2}}\\
\hline
System model & State & $\text{M}_1$ & $\text{I}_1$ & $\text{Z}_1$ & $\text{C}_1$ & $\text{M}_2$ & $\text{I}_2$ & $\text{Z}_2$ & $\text{C}_2$ & E & $\text{E}_2$ & D \\
\hline
\multirow{3}{7em}{Isolated SISO} & $q_1$ & 0 & 0 & 1 & 0 & 0 & 0 & 0 & 0 & 1 & 0 & 0\\ 
& $q_2$ & 0 & 1 & 0 & 0 & 0 & 0 & 0 & 0 & 1 & 0 & 0 \\ 
& $q_3$ & 1 & 0 & 0 & 0 & 0 & 0 & 0 & 0 & 0 & 0 & 0 \\ 
\hline
\multirow{4}{7em}{SISO with 1 downstream target} & $q_1$ & 0 & 0 & 0 & 1 & 0 & 0 & 0 & 0 & 1 & 0 & 0 \\ 
& $q_2$ & 0 & 0 & 1 & 0 & 0 & 0 & 0 & 0 & 1 & 0 & 1 \\ 
& $q_3$ & 0 & 1 & 0 & 0 & 0 & 0 & 0 & 0 & 1 & 0 & 1 \\
& $q_4$ & 1 & 0 & 0 & 0 & 0 & 0 & 0 & 0 & 0 & 0 & 1 \\
\hline
\multirow{8}{7em}{Isolated MIMO} & $q_1$ & 0 & 0 & 1 & 0 & 0 & 0 & 1 & 0 & 1 & 0 & 0 \\ 
& $q_2$ & 0 & 0 & 1 & 0 & 0 & 1 & 0 & 0 & 1 & 0 & 0 \\ 
& $q_3$ & 0 & 0 & 1 & 0 & 1 & 0 & 0 & 0 & 0 & 0 & 0 \\
& $q_4$ & 0 & 1 & 0 & 0 & 0 & 0 & 1 & 0 & 1 & 0 & 0 \\
& $q_5$ & 0 & 1 & 0 & 0 & 0 & 1 & 0 & 0 & 1 & 0 & 0 \\
& $q_6$ & 0 & 1 & 0 & 0 & 1 & 0 & 0 & 0 & 0 & 0 & 0 \\
& $q_7$ & 1 & 0 & 0 & 0 & 0 & 1 & 0 & 0 & 0 & 0 & 0 \\
& $q_8$ & 1 & 0 & 0 & 0 & 1 & 0 & 0 & 0 & 0 & 0 & 0 \\
\hline
\multirow{11}{7em}{MIMO with 1 downstream target} & $q_1$ & 0 & 0 & 0 & 1 & 0 & 0 & 1 & 0 & 1 & 0 & 0 \\ 
& $q_2$ & 0 & 0 & 0 & 1 & 0 & 1 & 0 & 0 & 1 & 0 & 0 \\ 
& $q_3$ & 0 & 0 & 0 & 1 & 1 & 0 & 0 & 0 & 0 & 0 & 0 \\
& $q_4$ & 0 & 0 & 1 & 0 & 0 & 0 & 1 & 0 & 1 & 0 & 1 \\
& $q_5$ & 0 & 0 & 1 & 0 & 0 & 1 & 0 & 0 & 1 & 0 & 1 \\
& $q_6$ & 0 & 0 & 1 & 0 & 1 & 0 & 0 & 0 & 0 & 0 & 1 \\
& $q_7$ & 0 & 1 & 0 & 0 & 0 & 0 & 1 & 0 & 1 & 0 & 1 \\
& $q_{8}$ & 0 & 1 & 0 & 0 & 0 & 1 & 0 & 0 & 1 & 0 & 1 \\
& $q_{9}$ & 0 & 1 & 0 & 0 & 1 & 0 & 0 & 0 & 0 & 0 & 1 \\
& $q_{10}$ & 1 & 0 & 0 & 0 & 0 & 0 & 1 & 0 & 0 & 0 & 1 \\
& $q_{11}$ & 1 & 0 & 0 & 0 & 0 & 1 & 0 & 0 & 0 & 0 & 1 \\
\hline
\multirow{15}{7em}{Two isolated SISO with MAC} & $q_1$ & 0 & 0 & 0 & 1 & 0 & 0 & 1 & 0 & 1 & 1 & 0 \\ 
& $q_2$ & 0 & 0 & 0 & 1 & 0 & 1 & 0 & 0 & 1 & 1 & 0 \\ 
& $q_3$ & 0 & 0 & 0 & 1 & 1 & 0 & 0 & 0 & 1 & 0 & 0 \\ 
& $q_4$ & 0 & 0 & 1 & 0 & 0 & 0 & 0 & 1 & 1 & 1 & 0 \\ 
& $q_5$ & 0 & 0 & 1 & 0 & 0 & 0 & 1 & 0 & 1 & 1 & 1 \\ 
& $q_6$ & 0 & 0 & 1 & 0 & 0 & 1 & 0 & 0 & 1 & 1 & 1 \\ 
& $q_7$ & 0 & 0 & 1 & 0 & 1 & 0 & 0 & 0 & 1 & 0 & 1 \\ 
& $q_8$ & 0 & 1 & 0 & 0 & 0 & 0 & 0 & 1 & 1 & 1 & 0 \\
& $q_9$ & 0 & 1 & 0 & 0 & 0 & 0 & 1 & 0 & 1 & 1 & 1 \\
& $q_{10}$ & 0 & 1 & 0 & 0 & 0 & 1 & 0 & 0 & 1 & 1 & 1 \\
& $q_{11}$ & 0 & 1 & 0 & 0 & 1 & 0 & 0 & 0 & 1 & 0 & 1 \\
& $q_{12}$ & 1 & 0 & 0 & 0 & 0 & 0 & 0 & 1 & 0 & 1 & 0 \\
& $q_{13}$ & 1 & 0 & 0 & 0 & 0 & 0 & 1 & 0 & 0 & 1 & 1 \\
& $q_{14}$ & 1 & 0 & 0 & 0 & 0 & 1 & 0 & 0 & 0 & 1 & 1 \\
& $q_{15}$ & 1 & 0 & 0 & 0 & 1 & 0 & 0 & 0 & 0 & 0 & 1 \\
\hline
\end{longtable}
\end{center}

\begin{figure}
    \centering
    \includegraphics[width=.4\linewidth]{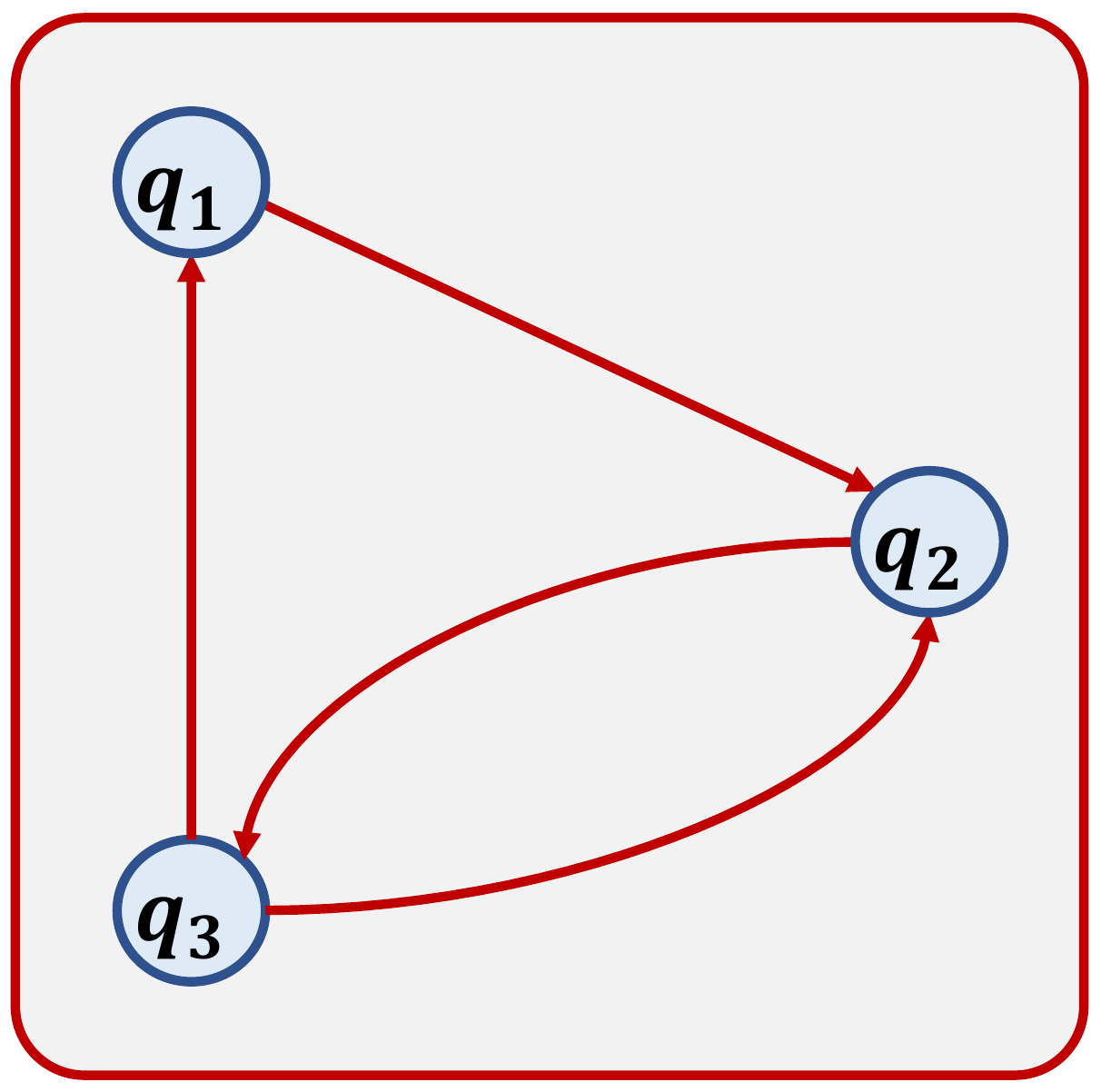}
    \caption{Transitions between microstates for the isolated SISO system model.}
    \label{fig: qtransitions}
\end{figure}
From now on, we report the mathematical steps only for the isolated SISO system, analogous results are obtained for the remaining models. Both in the case of $\text{I}_{\text{tot}_1} = 0$ and of $\text{I}_{\text{tot}_1} = 1$, the set of chemical reactions representing the biomolecular system can be rewritten as a set of transitions between the enumerated microstates, as depicted in Fig.~\ref{fig: qtransitions} for the isolated SISO system in the case of $\text{I}_{\text{tot}_1} = 1$. Clearly, since the case with $\text{I}_{\text{tot}_1} = 0$ is characterized only by one microstate, the system remains in $q_1$ with probability equal to 1 in each instant of time $t$, i.e., $P\left(q_1,t\right) = 1$. When $\text{I}_{\text{tot}_1} = 1$, and assuming that all reactions take place in a volume $\Omega$, it is possible to use the propensity functions, i.e., the probability that a given reaction occurs in a sufficiently small amount of time, to obtain
\begin{equation}
\begin{array}{lcc}
    \ce{$\text{I}_1$ + E ->[$\text{k}_0^+$] $\text{M}_1$}: \quad q_2 \to q_3; \quad a_1(q_2) = \textbf{$\text{k}_0^+$} = \left(\nicefrac{\text{k}_0^+}{\Omega}\right)\\
    \ce{$\text{I}_1$ + E <-[$\text{k}_0^-$] $\text{M}_1$}: \quad q_3 \to q_2; \quad a_2(q_3) = \text{k}_0^-\\
    \ce{$\text{M}_1$ ->[$\text{c}_1$] E + $\text{Z}_1$}: \quad q_3 \to q_1; \quad a_3(q_3) = \text{c}_1\\
    \ce{$\text{Z}_1$ ->[$\text{c}_2$] $\text{I}_1$}: \quad q_1 \to q_2; \quad a_4(q_1) = \text{c}_2\nonumber.
\end{array}
\end{equation}
Then, the CME can be written down using the propensity functions for each reaction. For the isolated SISO system it becomes
\begin{equation}
    \frac{d}{dt}\begin{bmatrix}P\left(q_1,t\right)\\ P\left(q_2,t\right)\\ P\left(q_3,t\right)\end{bmatrix} = \begin{bmatrix} -\text{c}_2 & 0 & \text{c}_1 \\ \text{c}_2 &  -\textbf{$\text{k}_0^+$} & \text{k}_0^- \\ 0 &  \textbf{$\text{k}_0^+$} & -\text{k}_0^- -\text{c}_1 \end{bmatrix}\begin{bmatrix}P\left(q_1,t\right)\\ P\left(q_2,t\right)\\ P\left(q_3,t\right)\end{bmatrix}.
\end{equation}
Then, the steady state solution for the probabilities $P\left(q_i,t_s\right)$ can be solved by setting $\dot P = 0$, which yields
\begin{align}
    &P\left(q_1,t_s\right) = \frac{\textbf{$\text{k}_0^+$}\text{c}_1}{\textbf{$\text{k}_0^+$}\text{c}_1+\text{c}_2\left(\text{k}_0^-+\textbf{$\text{k}_0^+$}+\text{c}_1\right)}\nonumber\\
    &P\left(q_2,t_s\right) = \frac{\text{c}_2\left(\text{k}_0^-+\text{c}_1\right)}{\textbf{$\text{k}_0^+$}\text{c}_1+\text{c}_2\left(\text{k}_0^-+\textbf{$\text{k}_0^+$}+\text{c}_1\right)} \\
    &P\left(q_3,t_s\right) = \frac{\textbf{$\text{k}_0^+$}\text{c}_2}{\textbf{$\text{k}_0^+$}\text{c}_1+\text{c}_2\left(\text{k}_0^-+\textbf{$\text{k}_0^+$}+\text{c}_1\right)}\nonumber.
\end{align}
Note that these probabilities can be seen as the joint pmf $P\left(\text{I}_1\left(t_s\right),\text{Z}_1\left(t_s\right)\right)$, where
\begin{align}\label{formula: ptsI1}
    &P\left(q_1,t_s\right) = P\left(\text{I}_1\left(t_s\right) = 0,\text{Z}_1\left(t_s\right) = 1\right),\nonumber\\
    &P\left(q_2,t_s\right) = P\left(\text{I}_1\left(t_s\right) = 1,\text{Z}_1\left(t_s\right) = 0\right),\\
    & P\left(q_3,t_s\right) = P\left(\text{I}_1\left(t_s\right) = 0,\text{Z}_1\left(t_s\right) = 0\right)\nonumber.
\end{align}
Thus, if we multiply $P\left(q_1,t\right) = 1$ by $P_{0_1}$ and~\eqref{formula: ptsI1} by $P_{1_1}$, we easily obtain the joint probability $P\left(\text{I}_1\left(t_s\right),\text{Z}_1\left(t_s\right), \text{I}_1\left(t_0\right)\right)$. By marginalizing over $\text{I}_1\left(t_s\right)$, we finally obtain 
\begin{equation}\label{formula: pjoint_extended}
    P\left(\text{I}_1\left(t_0\right),\text{Z}_1\left(t_s\right)\right) = \begin{bmatrix}P_{0_1} & 0\\
    \frac{\text{c}_2\left(\text{k}_0^-+\textbf{$\text{k}_0^+$}+\text{c}_1\right)}{\textbf{$\text{k}_0^+$}\text{c}_1+\text{c}_2\left(\text{k}_0^-+\textbf{$\text{k}_0^+$}+\text{c}_1\right)}P_{1_1} & \left(1-\frac{\text{c}_2\left(\text{k}_0^-+\textbf{$\text{k}_0^+$}+\text{c}_1\right)}{\textbf{$\text{k}_0^+$}\text{c}_1+\text{c}_2\left(\text{k}_0^-+\textbf{$\text{k}_0^+$}+\text{c}_1\right)}\right)P_{1_1}\end{bmatrix},
\end{equation}
where the rows represent the possible values assumed by $\text{I}_1\left(t_0\right)$, i.e., 0 and 1, and the columns the ones assumed by $\text{Z}_1\left(t_s\right)$.

By remembering that the dissociation constant is defined as $\text{k}_0 = \nicefrac{\text{k}_0^-}{\text{k}_0^+}$, we introduce $\textbf{k}_0 = \nicefrac{\text{k}_0^-}{\textbf{k}_0^+}$. Furthermore, we hypothesize $\text{c}_1, \text{c}_2 \ll \text{k}_0^-, \text{k}_0^+$. Then, we rewrite~\eqref{formula: pjoint_extended} as
\begin{equation}
    P\left(\text{I}_1\left(t_0\right),\text{Z}_1\left(t_s\right)\right) = \begin{bmatrix}P_{0_1} & 0\\
    \frac{\left(1 + \textbf{$\text{k}_0$}\right)\text{c}_2}{\text{c}_1+\left(1+ \textbf{$\text{k}_0$}\right)\text{c}_2}P_{1_1} & \left(1-\frac{\left(1 + \textbf{$\text{k}_0$}\right)\text{c}_2}{\text{c}_1+\left(1+ \textbf{$\text{k}_0$}\right)\text{c}_2}\right)P_{1_1}\end{bmatrix},
\end{equation}
where $\frac{\left(1 + \textbf{$\text{k}_0$}\right)\text{c}_2}{\text{c}_1+\left(1+ \textbf{$\text{k}_0$}\right)\text{c}_2} = A$.

Having $P\left(\text{I}_1\left(t_0\right),\text{Z}_1\left(t_s\right)\right)$, it is straightforward to obtain $P\left(\text{Z}_1\left(t_s\right)\right)$ by marginalizing over $\text{I}_1\left(t_0\right)$, and then
\begin{equation}
    P\left(\text{I}_{1}\left(t_0\right) \mid \text{Z}_1\left(t_s\right)\right) = \frac{P\left(\text{I}_1\left(t_0\right),\text{Z}_1\left(t_s\right)\right)}{P\left(\text{Z}_1\left(t_s\right)\right)} = \begin{bmatrix} \frac{P_{0_1}}{P_{0_1}+AP_{1_1}} & 0\\
    \frac{AP_{1_1}}{P_{0_1}+AP_{1_1}} & 1\end{bmatrix}.
\end{equation}
The entropies are now easily evaluated 
\begin{equation}
    H\left(\text{I}_{1}\left(t_0\right)\right) = -\sum_{j=0}^{1} P\left(\text{I}_{1_j}\left(t_0\right)\right)\log\left( P\left(\text{I}_{1_j}\left(t_0\right)\right)\right) = \log_e\left(P_{0_1}^{\left(-P_{0_1}\right)}P_{1_1}^{\left(-P_{1_1}\right)}\right),
\end{equation}
and
\begin{align}
    H\left(\text{I}_{1}\left(t_0\right)\mid \text{Z}_1\left(t_s\right)\right) &= -\sum_{i=0}^{1}\sum_{j=0}^{1} P\left(\text{Z}_{1_i}\left(t_s\right)\right)P\left(\text{I}_{1_j}\left(t_0\right)\mid \text{Z}_{1_i}\left(t_s\right)\right)\log \left(P\left(\text{I}_{1_j}\left(t_0\right) \mid \text{Z}_{1_i}\left(t_s\right)\right)\right) =\nonumber\\ &\log_e\left(\left(\left(\frac{P_{0_1}}{P_{0_1}+AP_{1_1}}\right)^{\left(-P_{0_1}\right)}\right)\cdot\left(\left(\frac{AP_{1_1}}{P_{0_1}+AP_{1_1}}\right)^{\left(-AP_{1_1}\right)}\right)\right).
\end{align}
In turn, the MI is
\begin{align}
    &I\left(\text{I}_1\left(t_0\right), \text{Z}_1\left(t_s\right)\right) = H\left(\text{I}_{1}\left(t_0\right)\right) - H\left(\text{I}_{1}\left(t_0\right)\mid \text{Z}_1\left(t_s\right)\right) = \nonumber\\
    &  \log_e\left(P_{0_1}^{\left(-P_{0_1}\right)}P_{1_1}^{\left(-P_{1_1}\right)}\right) - \log_e\left(\left(\left(\frac{P_{0_1}}{P_{0_1}+AP_{1_1}}\right)^{\left(-P_{0_1}\right)}\right)\cdot\left(\left(\frac{AP_{1_1}}{P_{0_1}+AP_{1_1}}\right)^{\left(-AP_{1_1}\right)}\right)\right)
\end{align}
nat/symbol.

\section*{Acknowledgment}
The authors would like to thank Theodore Wu Grunberg and Yili Qian for their valuable comments. F.R. acknowledges the support of the ``Progetto Roberto Rocca'' Doctoral Fellowship.

\ifCLASSOPTIONcaptionsoff
  \newpage
\fi

\bibliographystyle{IEEEtranTCOM}
\bibliography{Bibliography.bib}

\end{document}